%% file: bondwire.tex
\documentclass[10pt,letterpaper,twocolumn,journal]{IEEEtran}
\usepackage{cite}
\usepackage[pdftex]{graphicx}
\usepackage{amsmath}
\usepackage{upgreek}
\usepackage{amssymb}
\usepackage{mathtools}
\usepackage[tight,nice]{units}
\usepackage{multirow}
\usepackage[caption=false,font=footnotesize]{subfig}
\usepackage{url}
\usepackage{hyperref}
\newcommand{\diff}{\mathrm{d}}
\newcommand{\bs}{\boldsymbol}
\newcommand{\trans}{^{\!\top}}
\newcommand{\argmin}[1]{\underset{#1}{\operatorname{arg}\,\operatorname{min}}\;}
\graphicspath{{figures/}}

\begin{document}
\title{An Extended Analytic Model for the Heating of Bondwires}
\author{David Duque, Tomas Gotthans,
        Renaud Gillon, Sebastian Sch\"{o}ps 
\thanks{D. Duque and S. Sch\"{o}ps are with the Institut f\"{u}r Theorie 
Elektromagnetischer Felder of the 
Technische Universit\"{a}t Darmstadt, Schlo{\ss}gartenstra{\ss}e 8, 64289 Darmstadt, Germany.}
\thanks{T. Gothans is with the Department of Electrical Engineering of
the Brno University of Technology, Brno, Czech Republic.}
\thanks{R. Gillon is with ON Semiconductor
Belgium b.v.b.a., Westerring 15, 9700 Oudenaarde, Belgium.}}

\maketitle

\begin{abstract}
We present an extended analytic formula for
the calculation of the temperature profile
along a bondwire embedded in a package.
The resulting closed formula is built
by coupling the heat transfer equations of the 
bondwire and the surrounding moulding 
compound by means of auxiliary variables
that stem from an \emph{ad-hoc} linearisation
and mediate the wire-mould thermal interaction.
The model, which corrects 
typical simplifications in previously
introduced analytic 
models, is also optimised against carefully 
taken experimental samples representing fusing 
events of bondwires within real packages.
\end{abstract}

\begin{IEEEkeywords}
Bondwires, heat equation, mould compound,
thermal conduction, thermal radiation, heat kernel, Green's function.    
\end{IEEEkeywords}

\IEEEpeerreviewmaketitle

\section{Introduction}

\IEEEPARstart{T}{he} ever tightening
specifications imposed on modern 
integrated circuits (ICs) by the
fast pacing semiconductor industry   
demand the manufacturing
of more energy-efficient chips which are
constantly smaller in size. These 
smart-power ICs must undergo and withstand
a whole range of electrostatic discharges (ESD),
automotive pulses drive and short-circuit tests
which require high currents during short times
flowing through 
the electric connections. 
Therefore, a good understanding of
the interplay of the parameters defining 
the time-to-failure in these tests is 
essential to realise cost-effective, 
robust and fault-tolerant designs. 

Among the techniques to establish electric 
connection between a chip and the lead frame
(pins) during device assembly, wire-bonding
stands up as the most cost-effective one
\cite{bwire_bib3,bwire_bib4,bwire_bib5,bwire_bib7,bwire_bib8}.
In wire-bonding, tiny and fine
gold, aluminium, or copper wires are used as 
electric path between the chip and its package.   
As miniaturisation of the chips becomes inevitable,
the required diameter of the bondwires must also 
decrease. Since the electric power to 
the chip must be supplied through these wires,
high current densities may occur that heat up 
the wires causing a substantial increase of their 
temperature. If the temperature exceeds a predefined
value, damage of the compound or bondwire melting 
are among the most common source of failure in 
IC devices \cite{bwire_bib5}.

From the afore-described scenario,
the need among package engineers 
for an accurate  
formula that enables the fast 
dimensioning of bondwires and predicts 
their safe operation range in a particular 
application is very important. Ideally, 
these calculations should be carried out 
expediently and must involve
all the parameters describing a package.
Heretofore, several simplified 
analytic formulas for the calculation 
of the heating of bondwires and thus the 
estimation of their current capacities
have already been published~\cite{bwire_bib3,bwire_bib4,bwire_bib5,bwire_bib7,bwire_bib8}.
However, most of these formulas do not take 
into account the temperature dependency of the 
wire parameters and introduced simplifications
upon the relevant heat transfer problem
that the resulting solution lacks all
the geometric information defining the 
package. In particular, in \cite{bwire_bib3}
by retaining the cylindrical symmetry of the
wire, the surrounding compound
is also deformed into a cylinder
in order to facilitate
coping with the heat boundary 
conditions along the 
wire-compound interface and this is done
despite disregarding heat flow 
along the cross section of the wire. 
At end, the model only accounts for 
radially outward heat conduction through
a cylindrical moulding compound of infinite 
extent and provides a loose  
treatment of the temperature
dependency of the wire parameters.
More recently in \cite{bwire_bib10},
the ideas presented in \cite{bwire_bib3} are expanded.
In particular, the moulding compound 
is still deformed into a material of 
cylindrical shape upon 
which on its outer surface either 
adiabatic or isothermal
boundary conditions are imposed.
In this manner, a finite cylindrical compound surrounding 
the wire is considered. The model employs
a coupling between the wire and compound 
heat equations and solves them numerically providing
thus the flexibility of analysing current pulses
of arbitrary shape. As before, only radial outward
heat conduction is accounted for.
and the geometric information of the 
package is again lost. 

In this paper, we address these typical
shortcomings by developing yet 
an analytic formulation
for the determination of the temperature
in bondwires. The more robust analytic
formula does involve the essential
physical parameters that define the
package, i.e., moulding compound
material and dimensions, bond-wire characteristics, etc.,
by using an appropriate set of heat transfer boundary
conditions (BCs) and constructing the 
heat kernel of the compound section.
The model couples the wire and compound heat
equations and mediates the interaction between 
them. Subsequent validation of
the model is performed and optimisation
with experimental data is also carried out. 

This paper is organised as follows:
In Section II, we formulate the heat
transfer problem at hand together
with the relevant BCs.
In Section III, we provide the solutions 
to the stated thermal system and perform
several numerical tests to 
validate them. In Section IV, we describe 
the procedure for optimising the model 
against experimental data, and also
how this data is acquired. This
section ends with a comparison on the perfomance 
of the model before and after optimisation.
Finally, in Section V, conclusions are given.
\section{Problem Formulation}
\begin{figure}[!htbp]
\centering
\includegraphics[width=0.550\columnwidth]{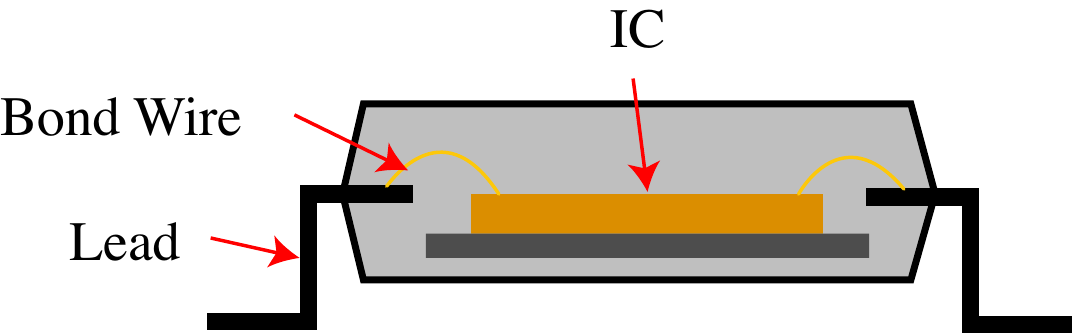}
\caption{Diagram of a classic IC lead-frame package.} 
\label{bwire_fig1}
\end{figure}

A simple diagram of a
classic IC lead-frame
package is depicted in Fig.~\ref{bwire_fig1}. 
 Because of the often
complicate arrangement of
the conductors within the package, 
simplifications are needed to formulate 
the relevant analytic heat transfer problem.
\begin{figure}
\centering
\includegraphics[width=0.80\columnwidth]{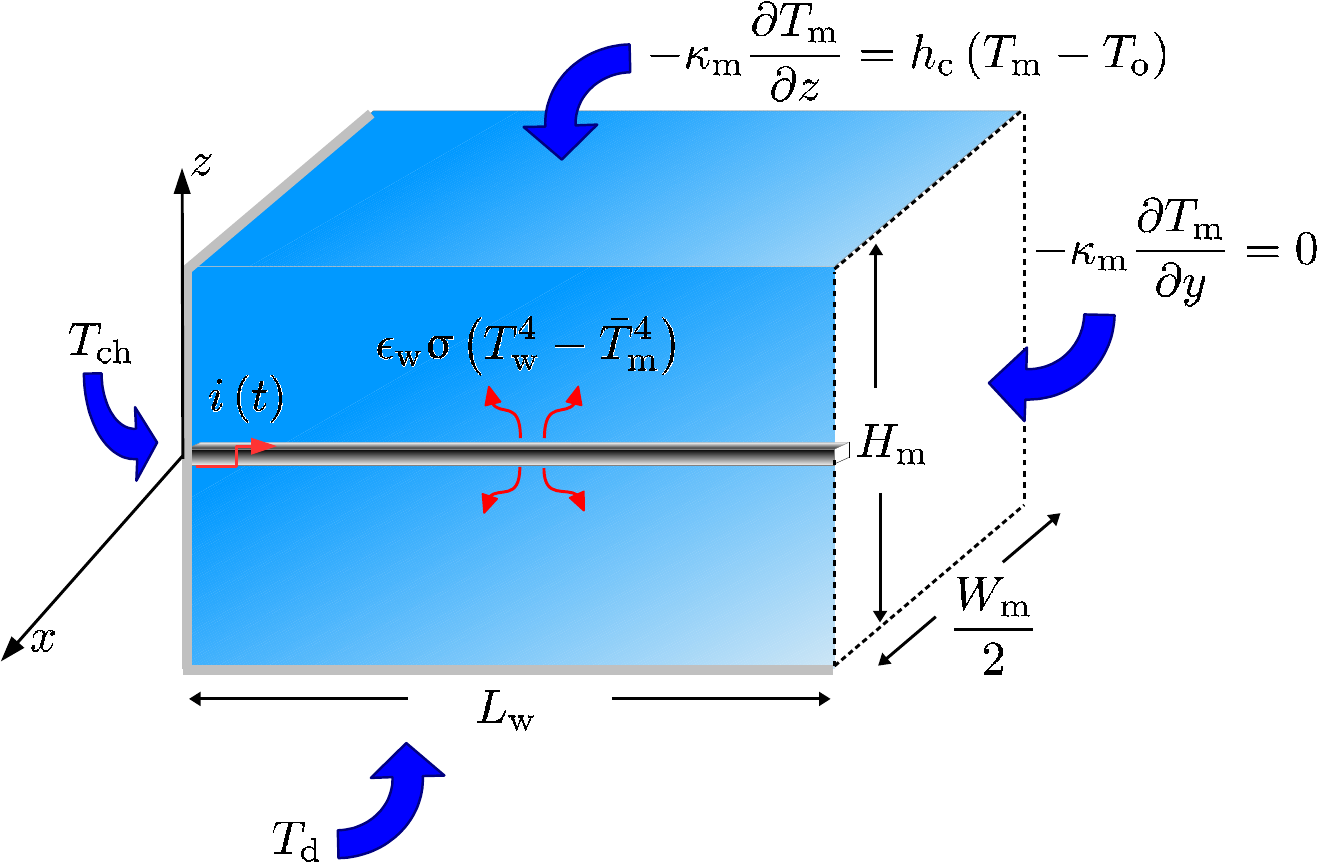}
\caption{Alternative bondwire heat transfer problem.}
\label{bwire_fig2}
\end{figure} 

In Fig.~\ref{bwire_fig2} we depict 
a simplified parametrical   
bondwire heat transfer problem. This problem consists of
the rectangular compound of height $H_{\text{m}}$ and 
width $W_{\text{m}}$ that defines
the package. The compound
is characterised by a homogeneous and 
isotropic thermal conductivity
$\kappa_{\text{m}}$, specific heat $c_{\text{e};\text{m}}$, 
and mass density $\rho_{\text{m}}$ whose 
temperature dependence is neglected for simplicity.
Similarly, the bondwire of length
$L_{\text{w}}$ is characterised by 
a homogeneous and isotropic thermal
conductivity $\kappa_{\text{w}}$,
specific heat $c_{\text{e};\text{w}}$,
mass density $\rho_{\text{w}}$, and electric
resistivity $\uprho_{\text{e};\text{w}}$.
The temperature dependence of 
$\kappa_{\text{w}}$ and $\uprho_{\text{e};\text{w}}$ is 
taken into account in the calculation
of the temperature $T_{\text{w}}(x,y,z,t)$
along the wire, viz. 
\begin{alignat}{2}
\label{bwire_eq1}
\kappa_{\text{w}}(\widetilde{T}_{\text{w}})&:=\kappa_{\text{o}}\left(1+\alpha_{\kappa}\widetilde{T}_{\text{w}}\right),&
\quad
\uprho_{\text{e;w}}(\widetilde{T}_{\text{w}})&:=\uprho_{\text{e;0}}
\left(1+\alpha_{\uprho}\widetilde{T}_{\text{w}}\right),
\end{alignat} with $\widetilde{T}_{\text{w}}\equiv T_{\text{w}}-T_{\text{0}}$, 
$T_{\text{0}}$ the reference 
(ambient) temperature,
$\alpha_{\kappa}$ the temperature coefficient of the 
thermal conductivity, and
$\alpha_{\uprho}$ the temperature coefficient of
the electric resistivity. The bondwire will be
heated up, during a time $t_{\text{p}}$, 
by the action of an electric current $i(t)$.
We aim at determining the time evolution
of $T_{\text{w}}$.

To provide an analytic solution, 
we impose suitable boundary conditions (BCs)
on the domain of interest 
$\Upomega=[0,W_{\text{m}}]\times[0,L_{\text{w}}]\times[0,H_{\text{m}}]$.
Namely, on the rightmost wall, we require 
that the heat flux mainly
occurs through the wire itself
(see Fig.~\ref{bwire_fig2}); therefore,
an adiabatic condition is enforced upon 
this wall, while the wire still remains at the 
lead temperature $T_{\text{ld}}$, viz.
\begin{alignat}{2}
\label{bwire_eq2}
-\kappa_{\text{m}}\frac{\partial}{\partial y} T_{\text{m}}\left(x,L_{\text{w}},z,t\right)&=0,&\quad
T_{\text{w}}\left(x,L_{\text{w}},z,t\right)&=T_{\text{ld}}.
\end{alignat} Here, $T_{\text{m}}$ is the compound temperature. 
On the leftmost wall the IC or chip is adjacently located 
(see Fig.~\ref{bwire_fig1}); thus we assume a constant
temperature $T_{\text{ch}}$
because of the chip high thermal capacitance, viz. 
\begin{alignat}{2}
\label{bwire_eq3} 
T_{\text{m}}\left(x,0,z,t\right)&=T_{\text{ch}},&\quad
T_{\text{w}}\left(x,0,z,t\right)&=T_{\text{ch}}.
\end{alignat} On the lateral and top walls of 
the compound, we assume convective heat transfer
because of the background medium
\cite{bwire_bib9}, viz. 
\begin{align}
-\kappa_{\text{m}}\frac{\partial}{\partial z}
T_{\text{m}}\left(x,y,\frac{H_{\text{m}}}{2},t\right)&=
h_{\text{c}}\left(T_{\text{m}}(x,y,\frac{H_{\text{m}}}{2},t)-T_{0})\right),\nonumber\\
\label{bwire_eq4}
-\kappa_{\text{m}}\frac{\partial}{\partial x}
T_{\text{m}}\left(\pm\frac{W_{\text{m}}}{2},y,z,t\right)&=
h_{\text{c}}\left(T_{\text{m}}(\pm\frac{W_{\text{m}}}{2},y,z,t)-T_{0})\right),
\end{align} where $h_{\text{c}}$ is the convective
transfer coefficient \cite{bwire_bib9}. On the
bottom wall, we assume a constant temperature
$T_{\text{d}}$ imposed by the die-attach (see Fig.~\ref{bwire_fig2}), viz.
\begin{equation}
\label{bwire_eq5}
T_{\text{m}}\left(x,y,-\frac{H_{\text{m}}}{2},t\right)=T_{\text{d}}.
\end{equation} Finally on the
wire surface $S_{\text{w}}$ we assume that both 
thermal conductivity and thermal radiation take place.
The latter is given by the Stefan-Boltzmann law \cite{bwire_bib9}.
Therefore upon the wire-mould interface we state
\begin{equation}
\label{bwire_eq6}  
-\int\limits_{S_{\text{w}}}\kappa_{\text{w}}\nabla_{\text{T}}T_{\text{w}}\cdot\diff\bs{S}=
\text{TH}_{\text{c}}+\int\limits_{S_{\text{w}}}\epsilon_{\text{w}}\upsigma\left(T^{4}_{\text{w}}-T^{4}_{0}\right)\diff S,
\end{equation} where $\nabla_{\text{T}}T_{\text{w}}$ is 
the \emph{transverse} gradient of the wire temperature,
that is the gradient along the $xz$-plane (see Fig.~\ref{bwire_fig2}), 
TH$_{\text{c}}$ just denotes the conductive part 
of the thermal flux that occurs on the wire surface, 
$\epsilon_{\text{w}}$
is the wire emissivity \cite{bwire_bib9}, and $\upsigma$ is the 
Stefan-Boltzmann constant \cite{bwire_bib9}. 

\subsection{Bondwire Heat Equation}

The heat equation for a stationary 
wire of constant mass density 
reads \cite{bwire_bib9}
\begin{equation}
\label{bwire_eq7} 
\rho_{\text{w}}c_{\text{e;w}}\frac{\partial T_{\text{w}}}{\partial t}=
\nabla\cdot\left(\kappa_{\text{w}}\nabla T_{\text{w}}\right)+\dot{q}_{\text{i}}, 
\end{equation} where $\dot{q}_{\text{i}}$ is the
\emph{impressed} volumetric thermal power density within
the wire. By expanding the above gradient operator as 
$\nabla=\partial_{y}\bs{a}_{y}+\nabla_{\text{T}}$,
with $\bs{a}_{y}$ the unit vector along the $y$-axis,
and assuming that the Biot number
\cite{bwire_bib9} of the wire is small along the $xz$-plane
\footnote{This assumption can be made insofar as the transverse dimensions
are much smaller than the longitudinal dimension. A condition
that is easily satisfied by bondwires.}, that is
$T_{\text{w}}$ mainly
varies along the wire axis, 
we may express \eqref{bwire_eq7} by means of
\eqref{bwire_eq6} in a form that explicitly involves the 
radiation condition, viz. 
\begin{equation}
\label{bwire_eq9}
\rho_{\text{w}}c_{\text{e};\text{w}}\frac{\partial}{\partial t}T_{\text{w}}=
\frac{\partial}{\partial y}\left(\kappa_{\text{w}}\frac{\partial}{\partial y}
T_{\text{w}}\right)-\epsilon_{\text{w}}\upsigma\left(T^{4}_{\text{w}}
-T^{4}_{0}\right)\frac{C_{\text{w}}}{A_{\text{w}}}+\dot{q}_{\text{i}},
\end{equation} where $A_{\text{w}}(y)$ and $C_{\text{w}}(y)$
are the cross-section area and perimeter 
of the wire, respectively. In \eqref{bwire_eq9}
we may think as if the contribution of 
TH$_{\text{c}}$ is \emph{apparently} vanished.
However, as we will see in section \ref{bwire_sec3},
the contribution of TH$_{\text{c}}$
is again accounted for with the introduction of a
linearising constant in the wire and compound 
heat equations.

\subsection{Moulding Compound Heat Equation}

The moulding compound heat equation in integral form
reads \cite{bwire_bib9,bwire_bib11}
\begin{equation}
\label{bwire_eq11} 
\rho_{\text{m}}c_{\text{e;m}}
\frac{\partial}{\partial t}\int\limits_{V_{\text{m}}}T_{\text{m}}
\diff V=\kappa_{\text{m}}\int\limits_{S_{\text{w}}}
\nabla T_{\text{m}}\cdot\diff\bs{S}+
\kappa_{\text{m}}\int\limits_{\tilde{S}_{\text{m}}}
\nabla T_{\text{m}}\cdot\diff\bs{S},
\end{equation} where $V_{\text{m}}$ 
is the compound volume, $S_{\text{w}}$ is the 
wire-compound common interface, and 
$\tilde{S}_{\text{m}}$ is the compound 
remaining surface. With the help of \eqref{bwire_eq6}
and the low-Biot-number assumption, we may 
express \eqref{bwire_eq11} as follows
\begin{multline}
\label{bwire_eq12} 
\rho_{\text{m}}c_{\text{e;m}}
\frac{\partial}{\partial t}\int\limits_{V_{\text{m}}}T_{\text{m}}
\diff V=\kappa_{\text{m}}\int\limits_{\tilde{S}_{\text{m}}}
\nabla T_{\text{m}}\cdot\diff\bs{S}+\\
\int\limits_{V_{\text{w}}}\epsilon_{\text{w}}\upsigma
\left(T^{4}_{\text{w}}-T^{4}_{0}\right)
\frac{C_{\text{w}}}{A_{\text{w}}}\diff V,
\end{multline} with $V_{\text{w}}$ the 
wire volume. The above equation permits to
regard the problem at hand by one
in which the wire is considered infinitesimally  
thin. To this end, we take $\lim_{A_{\text{w}}\rightarrow 0}$ 
on \eqref{bwire_eq12}; thus, this can be written point-wise as follows  
\begin{equation}
\label{bwire_eq13} 
\rho_{\text{m}}c_{\text{e;m}}
\frac{\partial}{\partial t}T_{\text{m}}=\kappa_{\text{m}}\nabla^{2}T_{\text{m}}+
\epsilon_{\text{w}}\upsigma\left(T^{4}_{\text{w}}-T^{4}_{0}\right)C_{\text{w}}\delta(x)\delta(z),
\end{equation} where $\delta(.)$ is the Dirac delta \cite{bwire_bib12}.
Equation \eqref{bwire_eq13} is the
heat equation of the compound with the wire as an impressed heat source,
and its solution involves the heat kernel (Green's function)
of the compound~\cite{bwire_bib13}.

\section{Heat Transfer Problem Solution}
\label{bwire_sec3}
Equations \eqref{bwire_eq9} and \eqref{bwire_eq13}
constitute a non-linear coupled 
thermal system. In this section,
we carry out the solution of this system
by means of an \emph{ad-hoc}
linearisation.

\subsection{Bondwire Solution}

We start by linearising the radiation 
term in \eqref{bwire_eq9} as follows
\begin{equation}
\label{bwire_eq14}
\begin{split}
T^{4}_{\text{w}}-T^{4}_{0}&=
\left(T^{3}_{\text{w}}+T^{2}_{\text{w}}T_{0}+T_{\text{w}}T^{2}_{0}+T^{3}_{0}\right)
\left(T_{\text{w}}-T_{0}\right)\\
                                 &\approx\chi_{\text{w}}\left(T_{\text{w}}-T_{0}\right),
\end{split}
\end{equation} with $\chi_{\text{w}}$
a constant. This approximation is reasonable 
insofar as thermal radiation is not the heat transfer dominant term~\cite{bwire_bib9}. 
Therefore in this manner, with the introduction of
the constant $\chi_{\text{w}}$ we are globing in one 
both the contribution of conduction and radiation to 
the thermal flux at the wire surface.
Next, we employ the following transformation 
\begin{equation}
\label{bwire_eq15} 
\widetilde{\theta}_{\text{w}}\left(\widetilde{T}_{\text{w}}\right):=
\frac{1}{\kappa_{0}}\int\limits_{0}^{\widetilde{T}_{\text{w}}}
\kappa_{\text{w}}\left(s\right)\diff s, 
\end{equation} which implies that 
$\widetilde{\theta}_{\text{w}}=
\widetilde{T}_{\text{w}}+\alpha_{\kappa}/2\,\widetilde{T}^{2}_{\text{w}}$
and $\partial_{y}\widetilde{\theta}_{\text{w}}=
\kappa_{\text{w}}/\kappa_{0}\,\partial_{y}\widetilde{T}_{\text{w}}$; thus,
enabling us to write \eqref{bwire_eq9} as
\begin{equation}
\label{bwire_eq16}
\rho_{\text{w}}c_{\text{e};\text{w}}\frac{\partial}{\partial t}\widetilde{\theta}_{\text{w}}=
\kappa_{\text{o}}\frac{\partial^{2}}{\partial y^{2}}
\widetilde{\theta}_{\text{w}}-F_{\text{o;w;r}}\widetilde{\theta}_{\text{w}}
+G_{\text{o;w}}+\frac{1}{2}H_{\text{o;w;r}},
\end{equation} with
\begin{alignat*}{2}
G_{\text{o;w}}&=\frac{I^2_{0}\uprho_{\text{e;o}}}{A^{2}_{\text{w}}},&\quad
F_{\text{o;w;r}}&=\epsilon_{\text{w}}\upsigma\chi_{\text{w}}\frac{C_{\text{w}}}{A_{\text{w}}},
\end{alignat*}
\begin{equation}
\label{bwire_eq17}
H_{\text{o;w;r}}=\frac{2I^2_{0}\uprho_{\text{e;o}}\alpha_{\uprho}\widetilde{T}_{\text{w;e}}}
{A^{2}_{\text{w}}}+\epsilon_{\text{w}}\upsigma\chi_{\text{w}}\frac{C_{\text{w}}}{A_{\text{w}}}
\alpha_{\kappa}\widetilde{T}^{2}_{\text{w;e}}, 
\end{equation} and where we have approximated  
$\partial_{t}\widetilde{\theta}_{\text{w}}\approx
\partial_{t}\widetilde{T}_{\text{w}}$
so as to keep linear the transient term
in \eqref{bwire_eq16} and have introduced 
$\widetilde{T}_{\text{w;e}}$ as an 
\emph{auxiliary} constant dubbed the wire
\emph{effective} temperature.

We solve \eqref{bwire_eq16} by separation of variables with
$\widetilde{\theta}_{\text{w}}(y,t)=\tilde{\theta}_{\text{w};1}(y,t)+
\tilde{\theta}_{\text{w};2}(y)$, thus yielding
\begin{equation}
\label{bwire_eq18}
\begin{split}
\tilde{\theta}_{\text{w}}(y,t)&=\sum_{k}C^{\text{t}}_{\text{w};k;\text{r}}
e^{-\frac{\kappa_{\text{o}}}{\rho_{\text{w}}c_{\text{e;w}}}\lambda^{2}_{y;\text{w},k}t}
e^{-\frac{F_{\text{o;w;r}}}{\rho_{\text{w}}c_{\text{e;w}}}t}\sin\left(\lambda_{y;\text{w},k}y\right)\\
           &+C^{\text{s}}_{1;y;\text{w;r}}\cosh\left(\sqrt{\frac{F_{\text{o;w;r}}}{\kappa_{\text{o}}}}y\right)
           +C^{\text{s}}_{2;y;\text{w;r}}\sinh\left(\sqrt{\frac{F_{\text{o;w;r}}}{\kappa_{\text{o}}}}y\right)\\
           &+\frac{1}{2}\frac{H_{\text{o;w;r}}}{F_{\text{o;w;r}}}+\frac{G_{\text{o;w}}}{F_{\text{o;w;r}}};\;
           \lambda_{y;\text{w},k}=\frac{k\pi}{L_{\text{w}}},\,k>0. 
\end{split}
\end{equation} Above, the coefficients
$\lbrace C^{\text{t}}_{\text{w};k;\text{r}}, C^{\text{s}}_{1;y;\text{w;r}},
C^{\text{s}}_{2;y;\text{w;r}}\rbrace$ are determined by means of
the initial condition at $t=0$ and
the relevant BCs in \eqref{bwire_eq2} and \eqref{bwire_eq3}, 
respectively. We observe that the auxiliary constants $\widetilde{T}_{\text{w;e}}$
and $\chi_{\text{w}}$ in \eqref{bwire_eq17} are yet
to be determined.

\subsection{Moulding Compound Solution}

Let us rewrite \eqref{bwire_eq13}
with the help of $\chi_{\text{w}}$, viz.
\begin{equation}
\label{bwire_eq19} 
\rho_{\text{m}}c_{\text{e;m}}
\frac{\partial}{\partial t}T_{\text{m}}=\kappa_{\text{m}}\nabla^{2}T_{\text{m}}+
\epsilon_{\text{w}}\upsigma\chi_{\text{w}}\widetilde{T}_{\text{w}}C_{\text{w}}\delta(x)\delta(z).
\end{equation} We notice in \eqref{bwire_eq19}
that if the impressed source
were zero, the compound
temperature would be solely imposed by the
chip and die-attach temperatures
(see Fig.~\ref{bwire_fig2}). Once the electric
current is switched on, another temperature
component superimpose. Let us denote 
by $T_{\text{m}}$
the first of these components and solve for it  
by defining $\widetilde{T}_{\text{m}}\equiv T_{\text{m}}-T_{0}$, expanding   
$\widetilde{T}_{\text{m}}\left(x,y,z,t\right)=
\widetilde{T}_{\text{m};1}\left(x,y,z,t\right)+
\widetilde{T}_{\text{m};2;1}\left(x,y,z\right)+\widetilde{T}_{\text{m};2;2}\left(x,y,z\right)$,
and substituting in \eqref{bwire_eq19} together with BCs 
\eqref{bwire_eq2}--\eqref{bwire_eq5}. 
The second component is obtained by means 
of the compound heat kernel (Green's function). 
In Appendix~\ref{bwire_appA}, we briefly 
describe how these
component functions are exactly determined.

Having calculated the aforementioned temperature
components, the moulding compound temperature
can be expressed as 
\begin{multline*}
T_{\text{m}}\left(x,y,z,t\right)=T_{0}+\widetilde{T}_{\text{m};1}\left(x,y,z,t\right)+
\widetilde{T}_{\text{m};2;1}\left(x,y,z\right)+\\
\widetilde{T}_{\text{m};2;2}\left(x,y,z\right)+
\int\limits_{0}^{t}\int\limits_{0}^{L_{\text{w}}}
G_{\text{m}}\left(x,y,z,t-\tau,y^{\prime}\right)\dot{q}_{\text{i}}\left(y^{\prime},\tau\right)
\diff y^{\prime}\diff \tau,
\end{multline*}
\begin{equation}
\label{bwire_eq20}
\dot{q}_{\text{i}}\left(y^{\prime},\tau\right)=
\epsilon_{\text{w}}\upsigma\chi_{\text{w}}\widetilde{T}_{\text{w}}\left(y^{\prime},\tau\right)C_{\text{w}}. 
\end{equation} The above expression provides
the temperature at any point within the compound 
for any time-variant current. Both auxiliary constants
$\widetilde{T}_{\text{w;e}}$ and $\chi_{\text{w}}$ 
also appear in \eqref{bwire_eq20} within
$\widetilde{T}_{\text{w}}\left(y^{\prime},\tau\right)$.
In Appendix~\ref{bwire_appB},
we describe how these two constants are determined by 
mediating the thermal interaction between 
the compound and the wire.

\subsection{Numerical Test}

We have implemented the afore-described model 
and have  performed numerical tests for 
Gold ($\mathrm{Au}$), Copper ($\mathrm{Cu}$), and
Aluminium ($\mathrm{Al}$) wires of diameters
$D_{\text{w}}=\unit[\lbrace 0.8,1.0,\ldots,1.8,2.0\rbrace]{mil}$
and length $L_{\text{w}}=\unit[2.5]{mm}$.
A moulding compound with $W_{\text{m}}=\unit[4.45]{mm}$
and $H_{\text{m}}=\unit[1.48]{mm}$, and made of an Epoxy 
resin with $\kappa_{\text{m}}=\unit[0.870]{W/(m\cdot K)}$,
$c_{\text{e};\text{m}}=\unit[882]{J/(Kg\cdot K)}$, and 
$\rho_{\text{m}}=\unit[1860]{kg/m^{3}}$ is assumed.

To verify our solution,
we compute $T_{\text{m}}$ at the 
$xy$- and $yz$-planes (see Fig.~\ref{bwire_fig2})   
for an $\mathrm{Au}$-wire of $D_{\text{w}}=\unit[2]{mil}$,
which carries an electric current of $I_{0}=\unit[3.7]{A}$
during a time $t_{\text{p}}=\unit[500]{ms}$.
In concrete, we want verify 
the compound temperature expansion
coefficients (see Appendix~\ref{bwire_appA})
when satisfying the relevant BCs. We have also assumed herein 
that $T_{\text{ch}}=\unit[80]{^{\circ}C}$, $T_{\text{ld}}=\unit[40]{^{\circ}C}$,
$T_{\text{d}}=\unit[35]{^{\circ}C}$, $T_{0}=\unit[20]{^{\circ}C}$,
and $h_{\text{c}}=\unit[25]{W/(m^{2}\cdot K)}$.

\begin{figure}[htbp!]
\centering
\subfloat[\label{bwire_fig3a}]{\includegraphics[width=0.53\columnwidth]{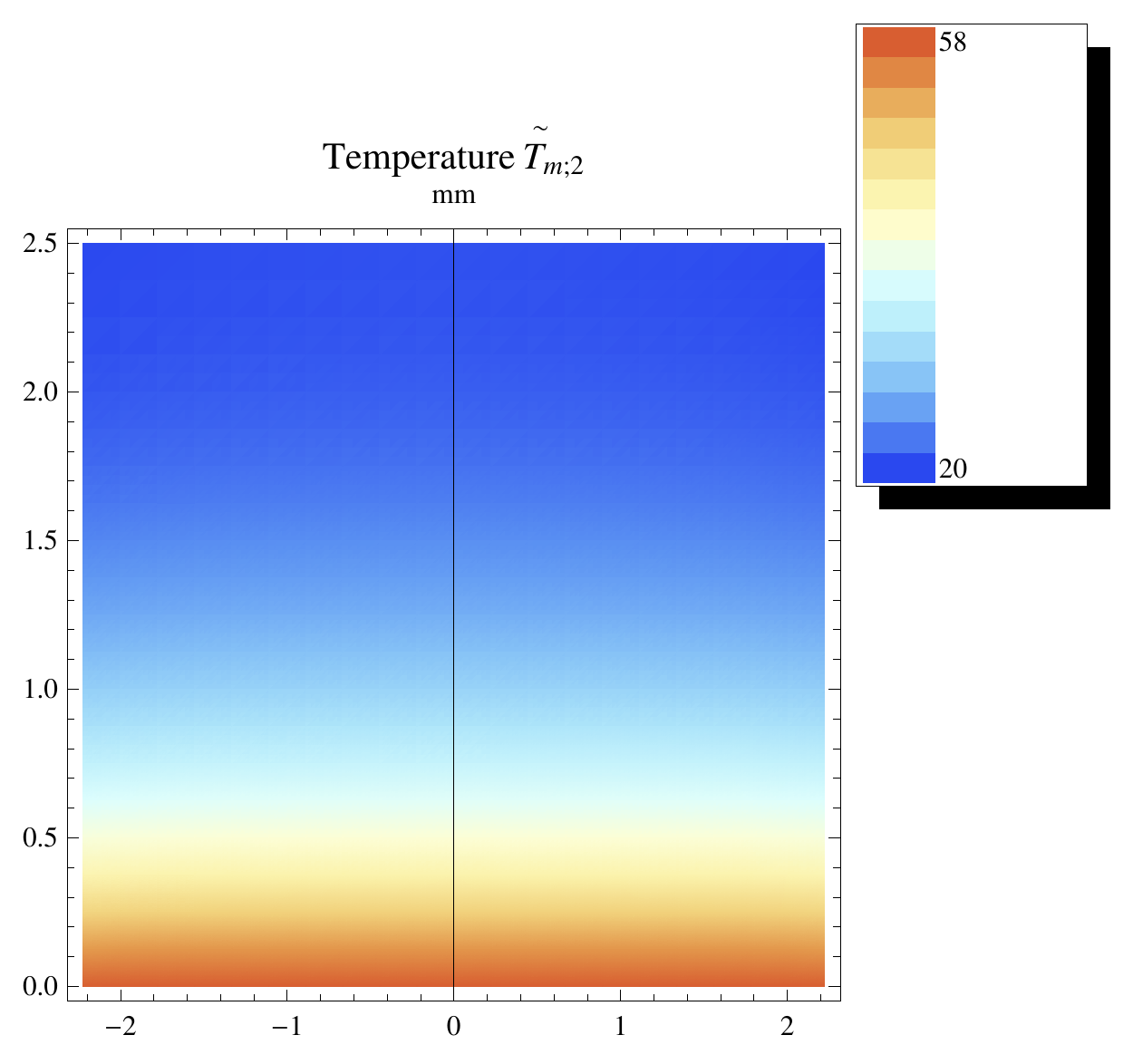}\hspace*{-0.1em}}
\subfloat[\label{bwire_fig3b}]{\includegraphics[width=0.53\columnwidth]{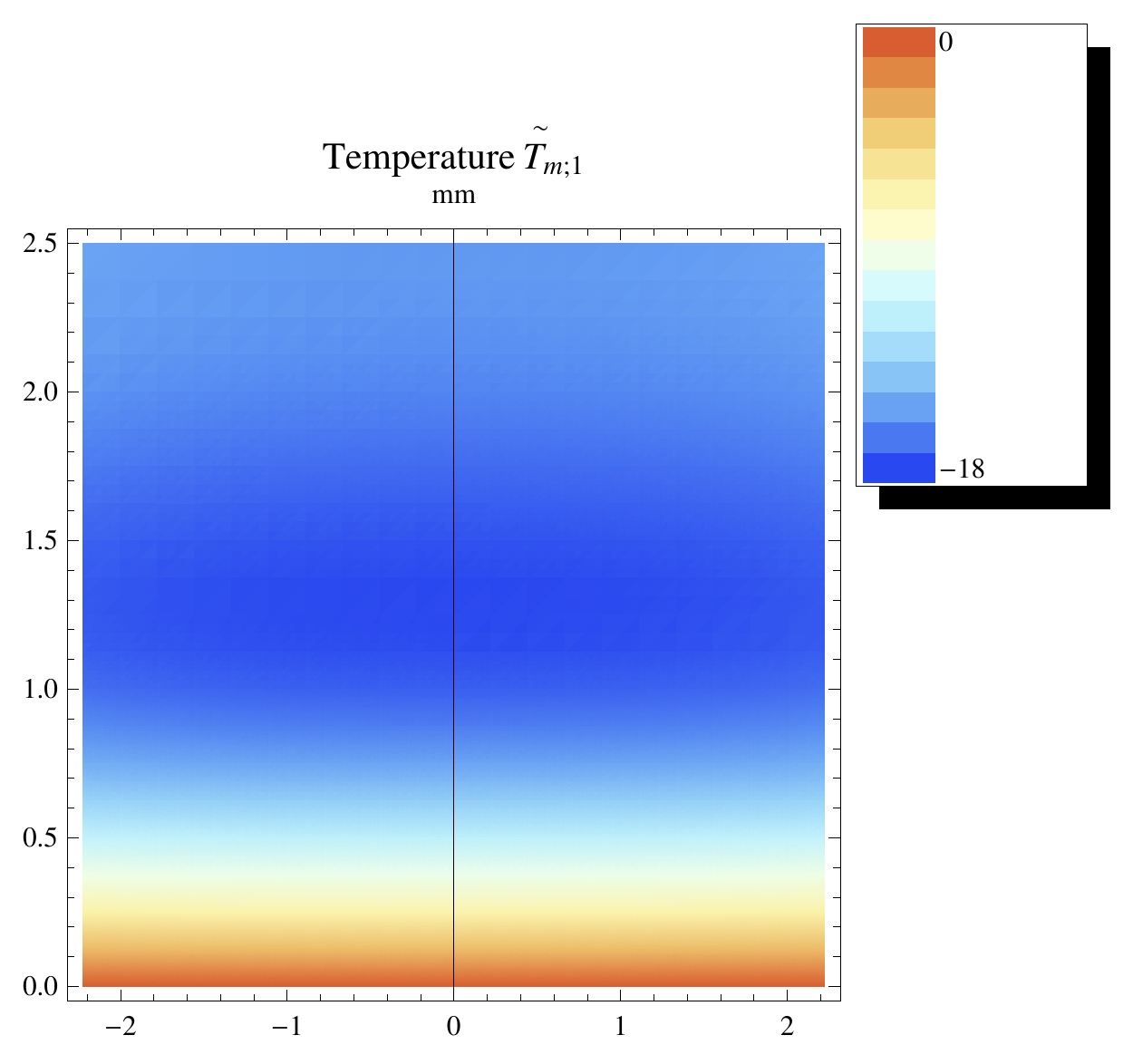}\hspace*{-0.1em}}\\
\subfloat[\label{bwire_fig3c}]{\includegraphics[width=0.53\columnwidth]{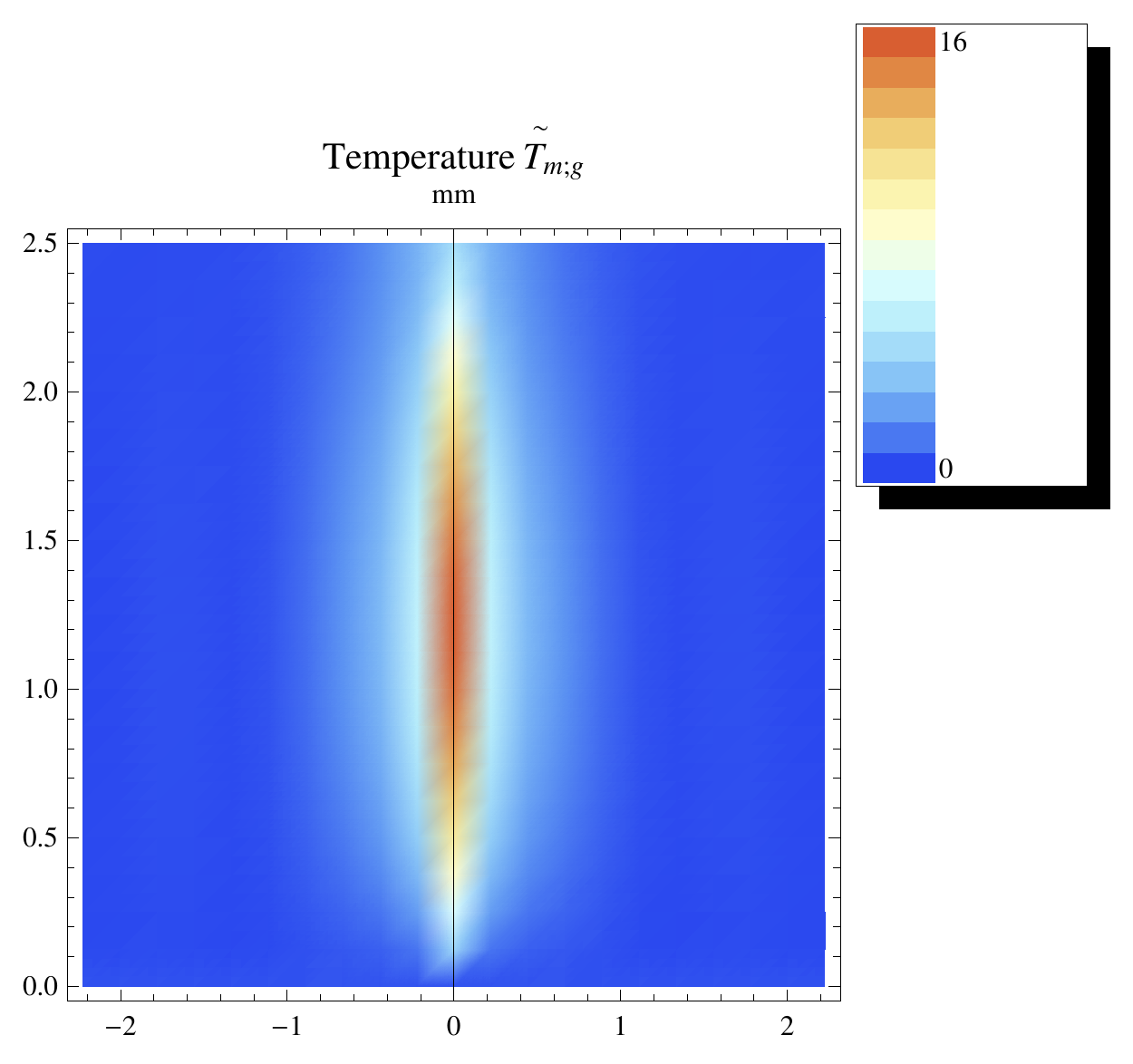}\hspace*{-0.1em}}
\subfloat[\label{bwire_fig3d}]{\includegraphics[width=0.53\columnwidth]{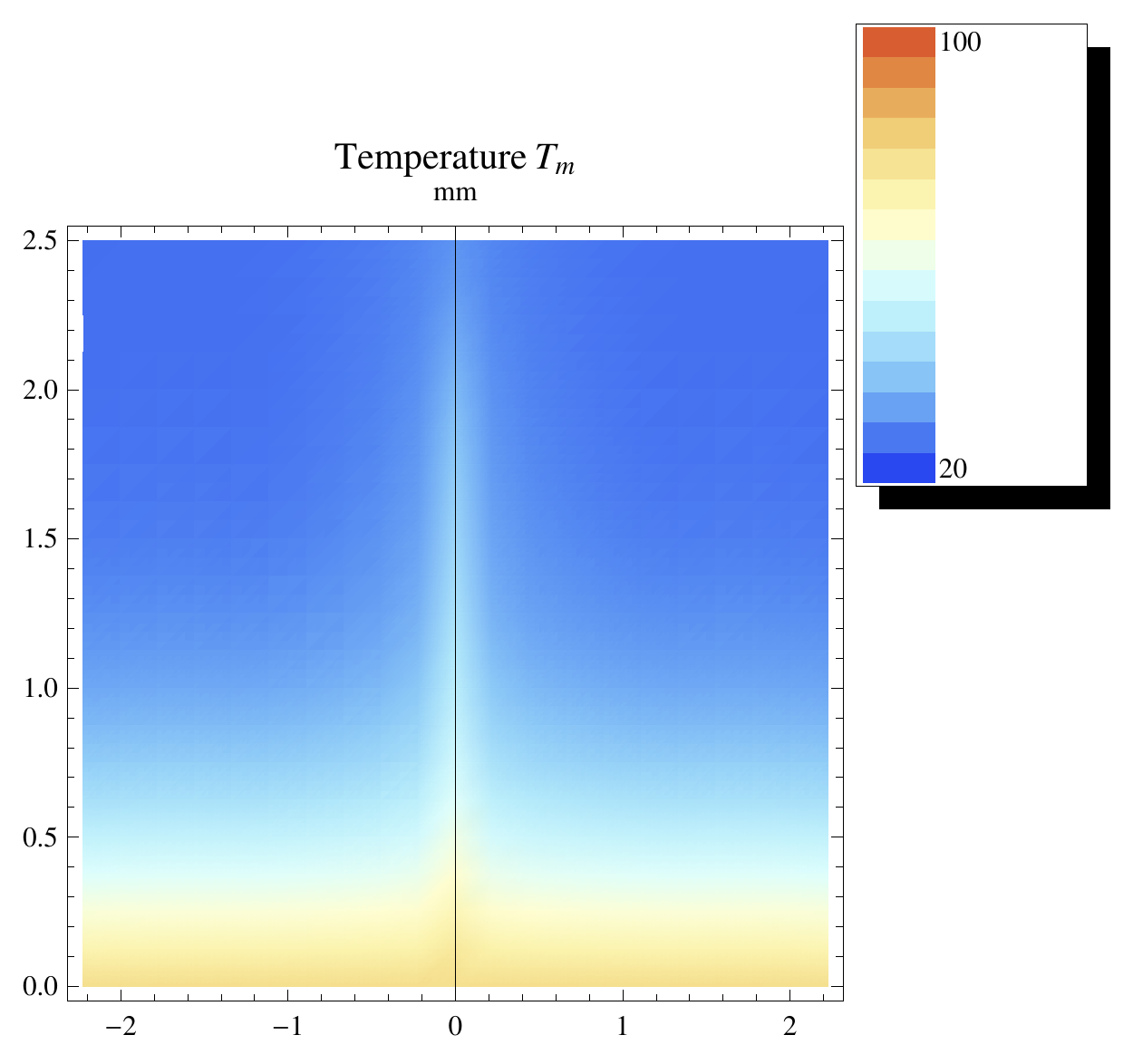}\vspace*{-0.2em}}\\
\caption{\label{bwire_fig3} Compound temperature components
at the $xy$-plane (top view) and $t=\unit[500]{ms}$;
(a) steady $\widetilde{T}_{\text{m};2}=\widetilde{T}_{\text{m};2,1}+\widetilde{T}_{\text{m};2,2}$ component;
(b) transient $\widetilde{T}_{\text{m};1}$ component; 
(c) heat kernel $\widetilde{T}_{\text{m;g}}$ component; (d) compound temperature $T_{\text{m}}$.
The $\mathrm{Au}$-wire is depicted as a straight line.}
\end{figure}

\begin{figure}[htbp!]
\centering
\subfloat[\label{bwire_fig4a}]{\includegraphics[width=0.515\columnwidth]{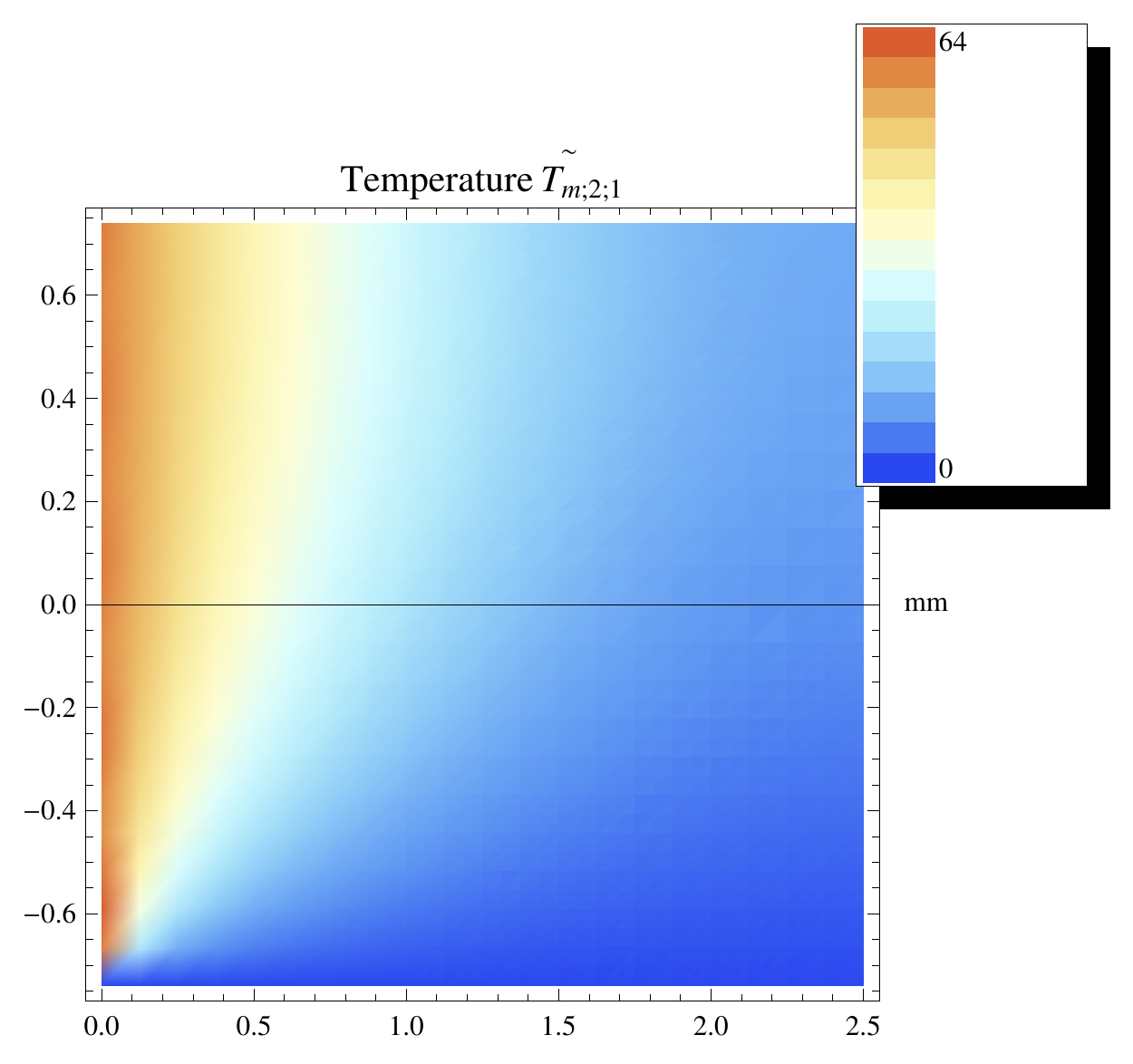}\hspace*{-0.1em}}
\subfloat[\label{bwire_fig4b}]{\includegraphics[width=0.515\columnwidth]{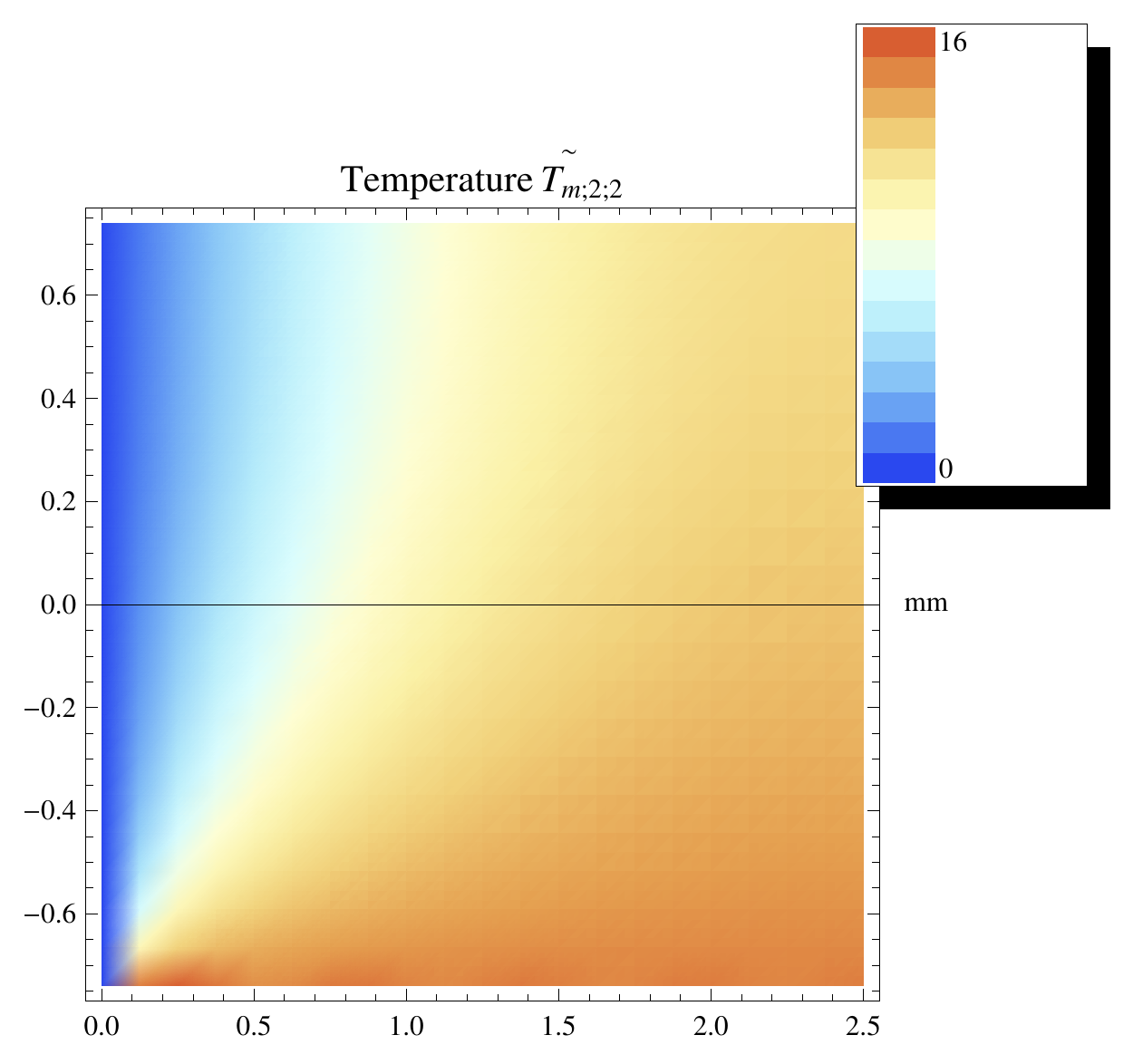}\hspace*{-0.1em}}\\
\subfloat[\label{bwire_fig4c}]{\includegraphics[width=0.515\columnwidth]{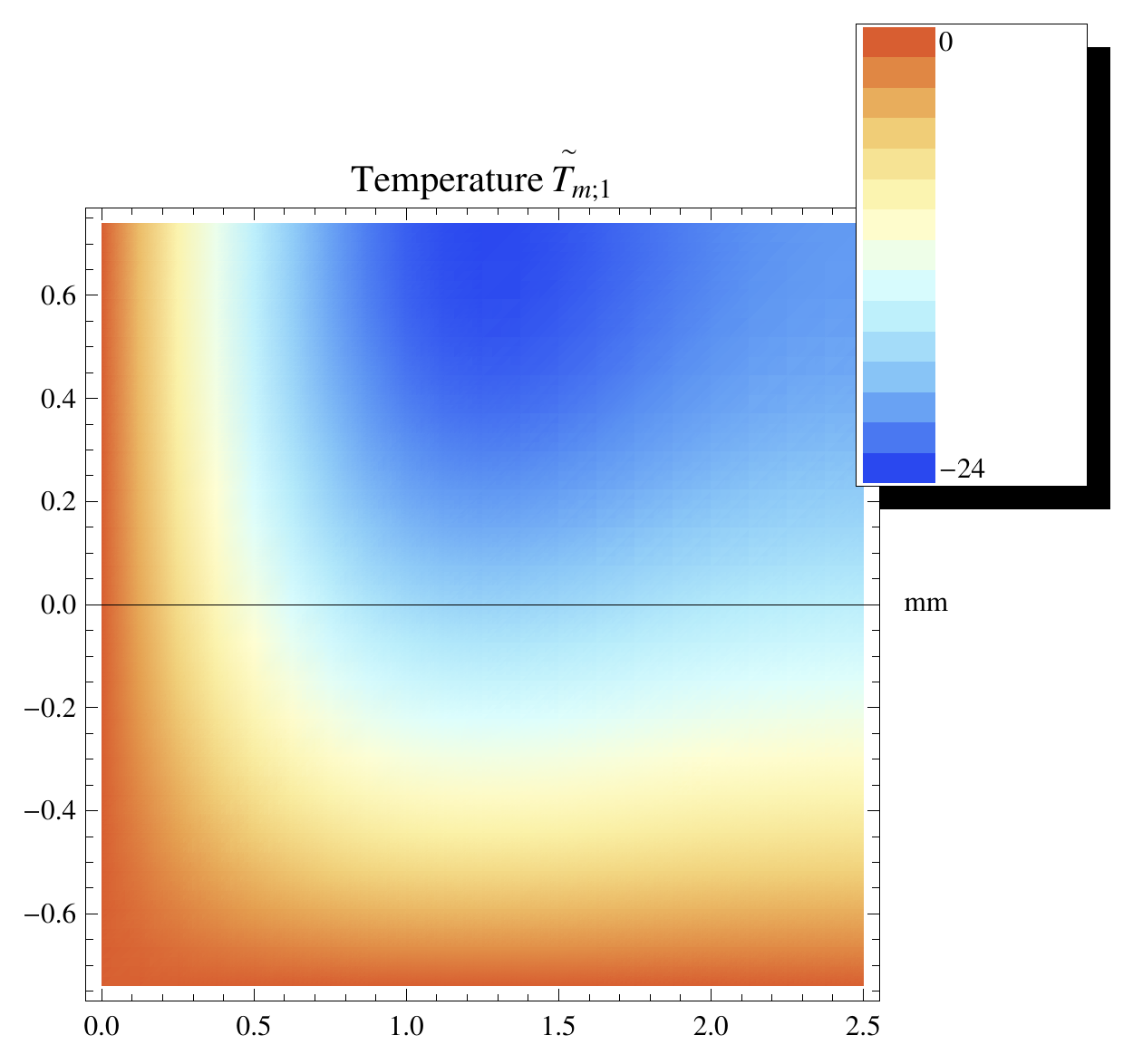}\hspace*{-0.1em}}
\subfloat[\label{bwire_fig4d}]{\includegraphics[width=0.515\columnwidth]{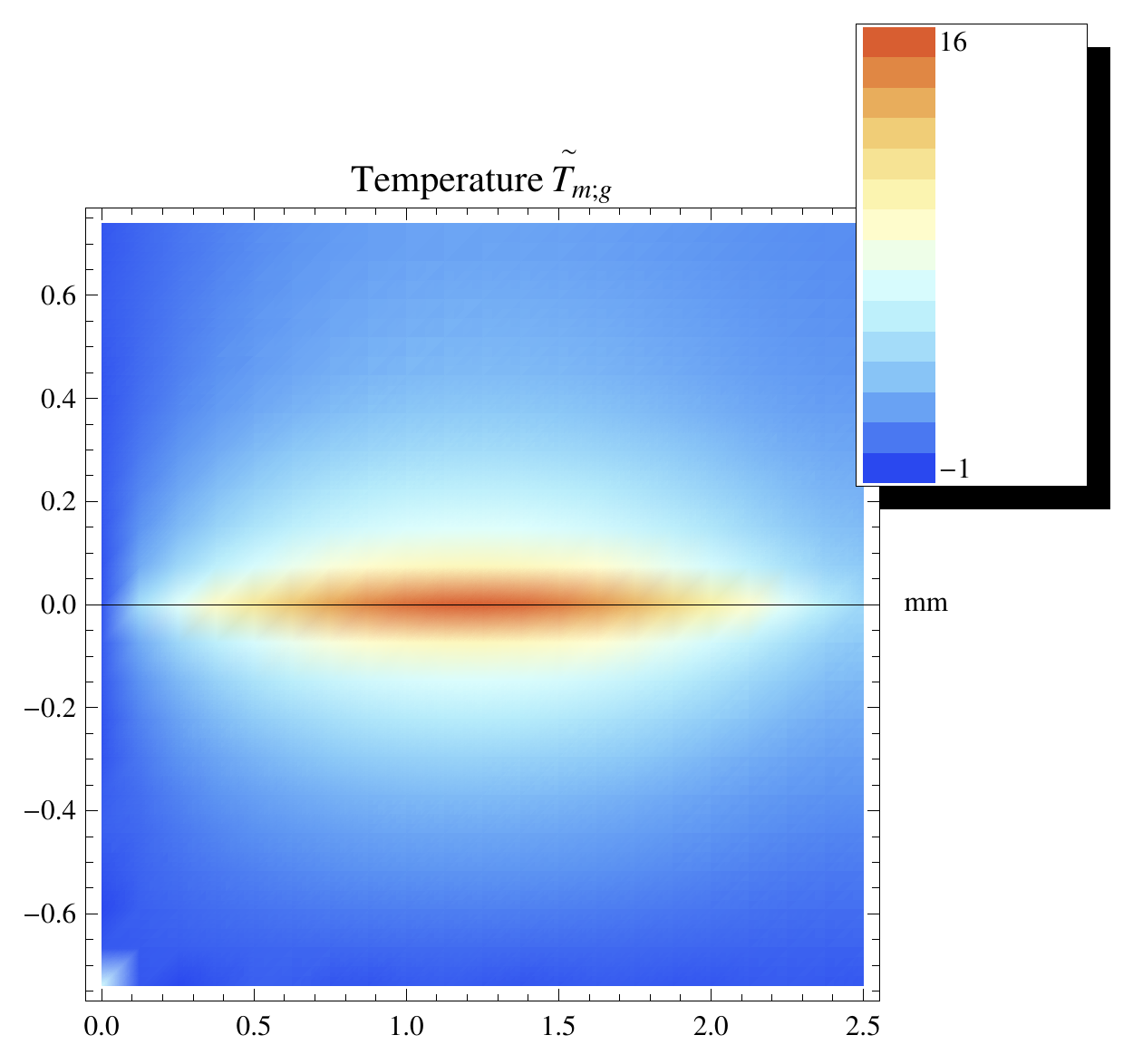}\hspace*{-0.1em}}\\
\subfloat[\label{bwire_fig4e}]{\includegraphics[width=0.515\columnwidth]{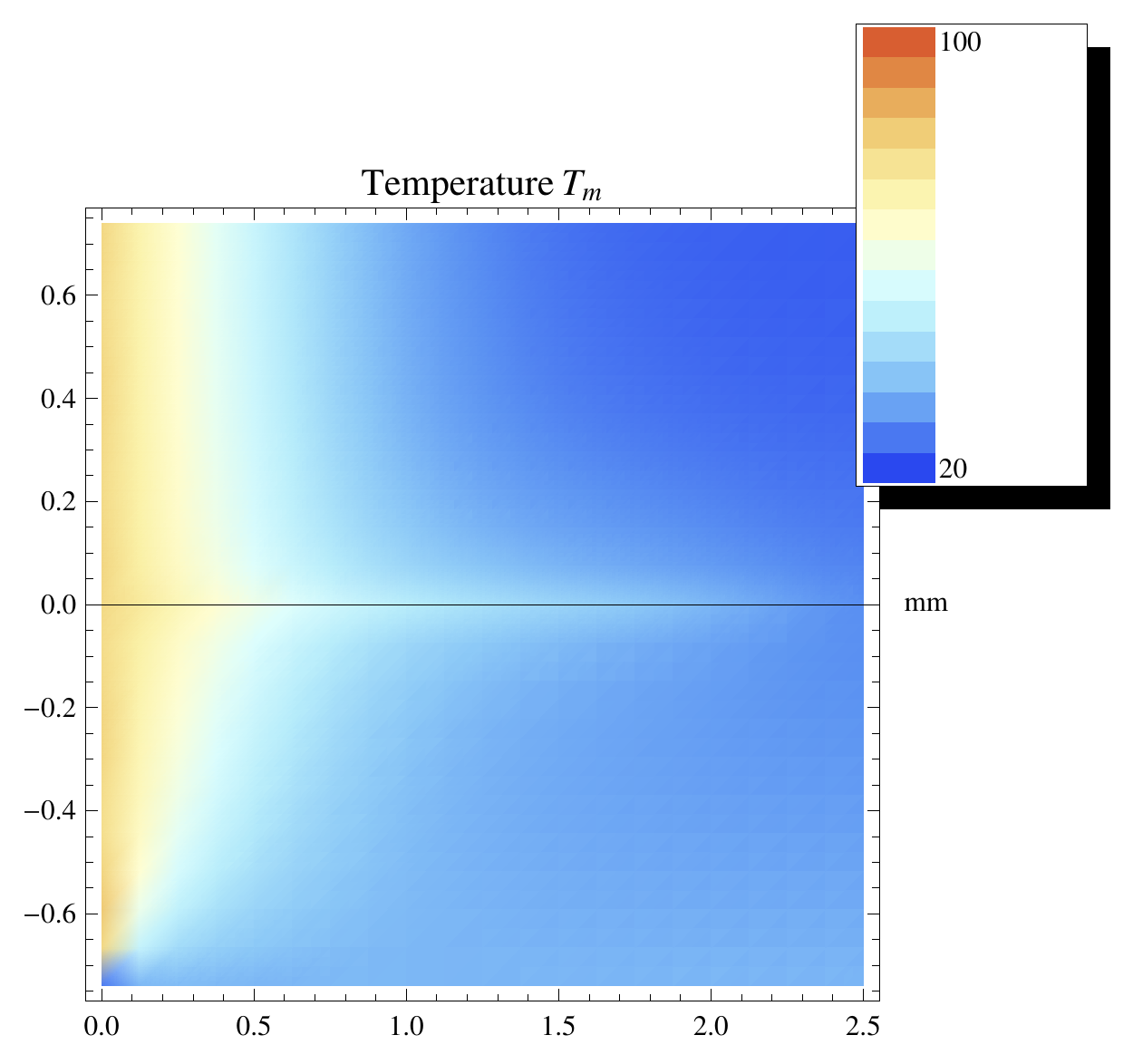}\vspace*{-0.2em}}\\
\caption{\label{bwire_fig4} Compound temperature components
at the $yz$-plane (side view) and $t=\unit[500]{ms}$;
(a) steady $\widetilde{T}_{\text{m};2,1}$ component; 
(b) steady $\widetilde{T}_{\text{m};2,2}$ component;
(c) transient $\widetilde{T}_{\text{m};1}$ component; 
(d) heat kernel $\widetilde{T}_{\text{m;g}}$ component,
(e) compound temperature $T_{\text{m}}$.
The $\mathrm{Au}$-wire is depicted as a straight line.}
\end{figure}

Fig.~\ref{bwire_fig3} and \ref{bwire_fig4} show
the plots of the temperature component in
\eqref{bwire_eq20} depicting the wire as a 
straight line. For example, Fig.~\ref{bwire_fig3a}
shows a top-view plot of the steady component
$\widetilde{T}_{\text{m};2}=
\widetilde{T}_{\text{m};2,1}+\widetilde{T}_{\text{m};2,2}$,
whereas Fig.~\ref{bwire_fig4a}
and \ref{bwire_fig4b} shows a side-view plot
of it separated into its constitutives 
$T_{\text{m};2,1}$ and $T_{\text{m};2,2}$.
In particular Fig.~\ref{bwire_fig3a}, \ref{bwire_fig4a},
and \ref{bwire_fig4b} show that the
steady component $\widetilde{T}_{\text{m};2}$
along the chip and die-attach planes  
satisfies the required BCs in~\eqref{bwire_eq24}
and \eqref{bwire_eq25}. Fig.~\ref{bwire_fig3b}
and \ref{bwire_fig4c} show the 
plot of the transient component $T_{\text{m};1}$,
and as we can note, this component
amounts to $\widetilde{T}_{\text{m};1}=0$
along the chip and die-attach planes,
thus verifying the required BCs in \eqref{bwire_eq24} and 
\eqref{bwire_eq25}.

Fig.~\ref{bwire_fig3c} and \ref{bwire_fig4d} shows the 
plot of the heat kernel component $\widetilde{T}_{\text{m;g}}$,
that is the integral in \eqref{bwire_eq20}.
These plots suggest that heat flux
propagates radially away from the wire into the 
compound and show a maximum that occurs at the 
wire mid-point while two minima appear towards
the extremes of the wire.
We also notice that this component is consistent 
with the required BCs along the chip and die-attach 
planes, where it vanishes. Finally in
Fig.~\ref{bwire_fig3d} and \ref{bwire_fig4e},
the combined action of all these components; namely,
the compound temperature $T_{\text{m}}$ is plotted.
In particular, we can notice therein that
$T_{\text{m}}=T_{\text{ch}}=\unit[80]{^{\circ}C}$
and $T_{\text{m}}=T_{\text{d}}=\unit[35]{^{\circ}C}$ 
along the chip and die-attach planes as demanded by 
the BCs. Furthermore, most of the compound is slightly 
above the (ambient) temperature $T_{0}=\unit[20]{^{\circ}C}$
used as the initial condition due to its low thermal 
conductivity. Nevertheless, we may expect
the compound to become hotter as time progresses.
It is important here to remark that $T_{\text{m}}$ is more an 
auxiliary scalar field that helps to estimate
the effect of the compound on the wire temperature,
rather than the \emph{true} compound temperature. 

\begin{figure}[htbp!]
\centering
\subfloat[$\mathrm{Al}$-wire]{\label{bwire_fig5a}
\includegraphics[width=0.52\columnwidth]{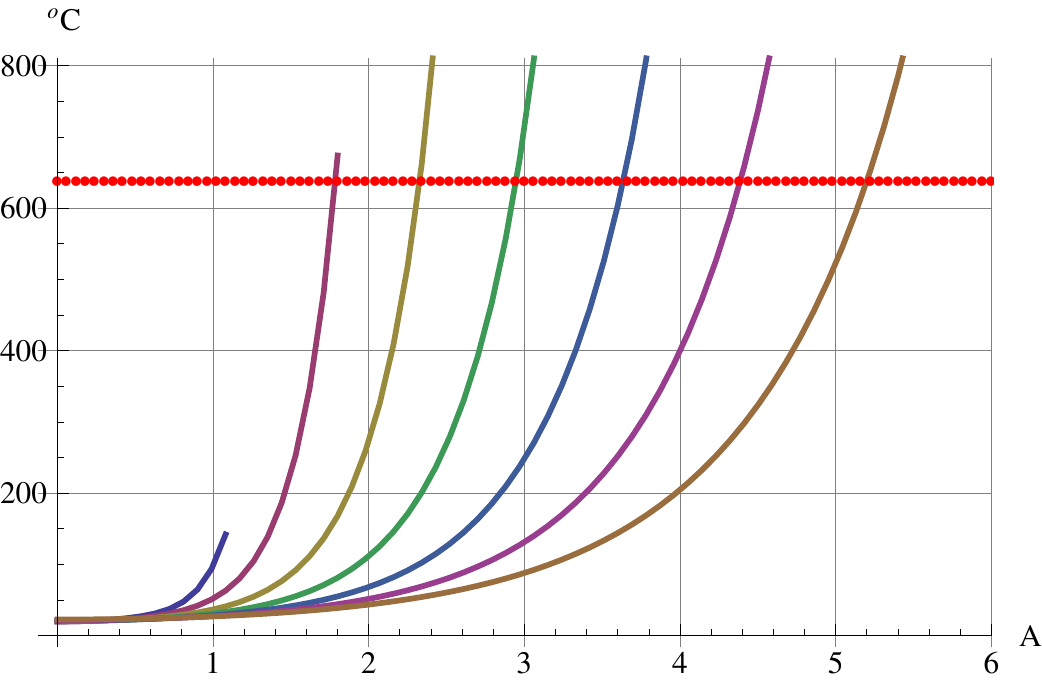}}
\subfloat[$\mathrm{Au}$-wire]{\label{bwire_fig5b}
\includegraphics[width=0.52\columnwidth]{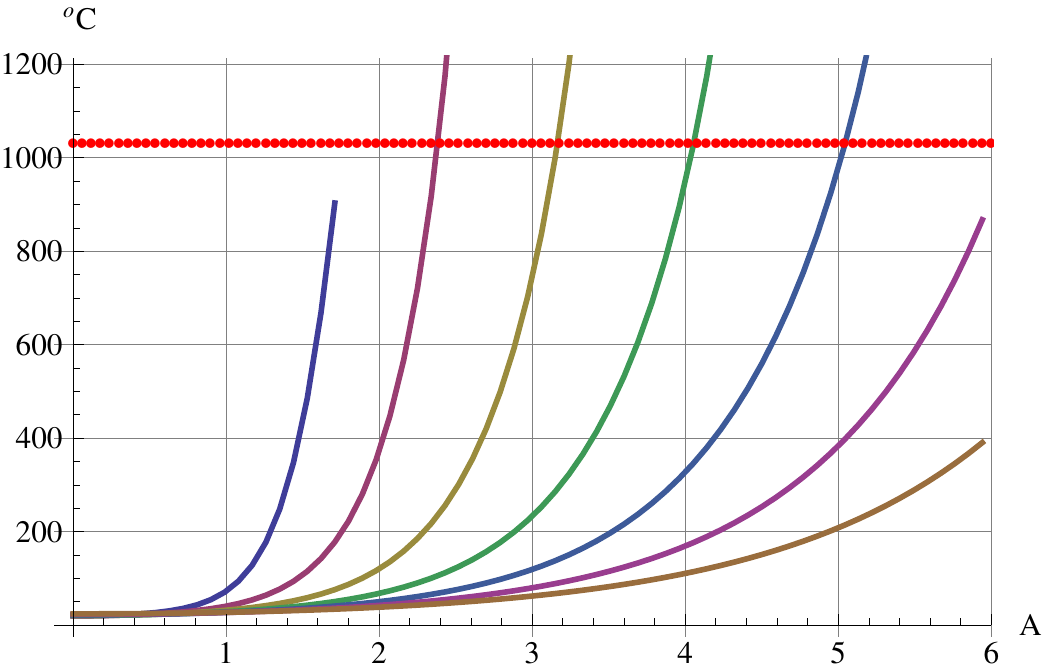}}\\
\subfloat[$\mathrm{Cu}$-wire]{\label{bwire_fig5c}
\includegraphics[width=0.52\columnwidth]{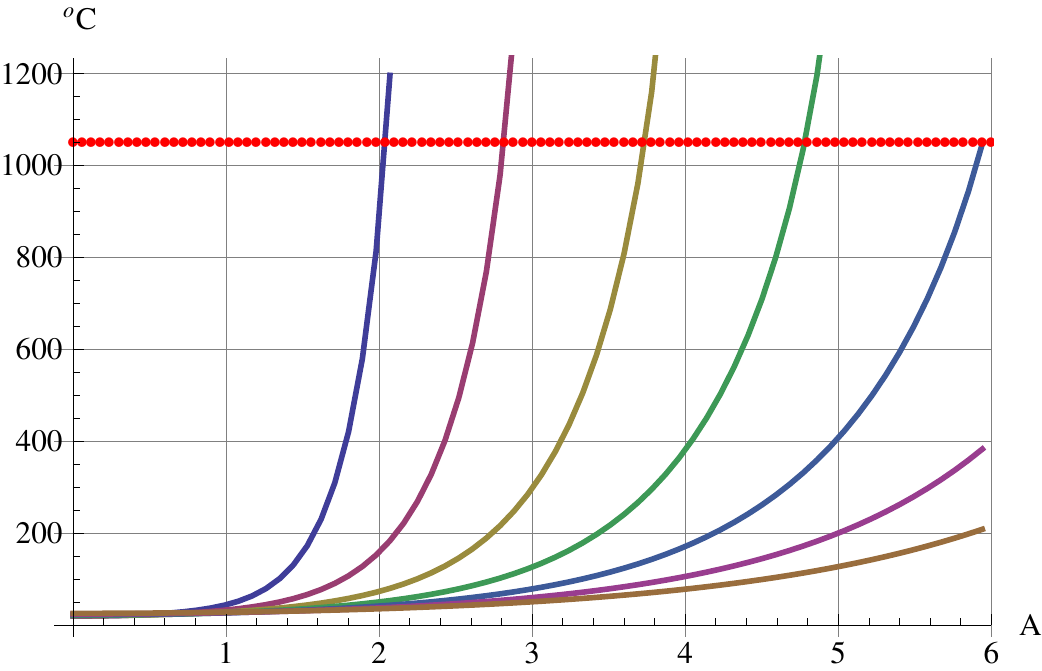}}\\
\caption{\label{bwire_fig5} Bondwire current capacities for
diameters $D_{\text{w}}=\unit[\lbrace 0.8,1.0,\ldots,1.8,2.0\rbrace]{mil}$ and
$L_{\text{w}}=\unit[2.5]{mm}$; a) $\mathrm{Al}$-wire; b) $\mathrm{Au}$-wire;
c) $\mathrm{Cu}$-wire. The horizontal dotted line indicates the
melting temperature.}
\end{figure}

Fig.~\ref{bwire_fig5} shows the estimated current
capacity $T_{\text{w}}$ \textit{vs} $I_{0}$
after a time of $\unit[50]{ms}$ for $\mathrm{Al}$-, $\mathrm{Au}$-, and $\mathrm{Cu}$-wires.
The temperature $T_{\text{w}}$ plotted therein
is at the wire mid-point. 
The results in Fig.~\ref{bwire_fig5}
reveals current capacities lower 
than those estimated in \cite{bwire_bib3} under similar setting. 
In other words, for a given bondwire configuration, 
the current amplitude in \cite{bwire_bib3} that causes
the wire to fuse after a certain time
is higher than the amplitude estimated by our model.  
This seems to indicate a tendency of the model 
in \cite{bwire_bib3} to underestimate the temperature in 
bondwires. Fig.~\ref{bwire_fig5} also 
demonstrates the capabilities of our model 
to provide a safe range of operation for the 
bondwires before melting and moulding 
deterioration is reached.

\begin{figure}
\centering
\includegraphics[width=0.70\columnwidth]{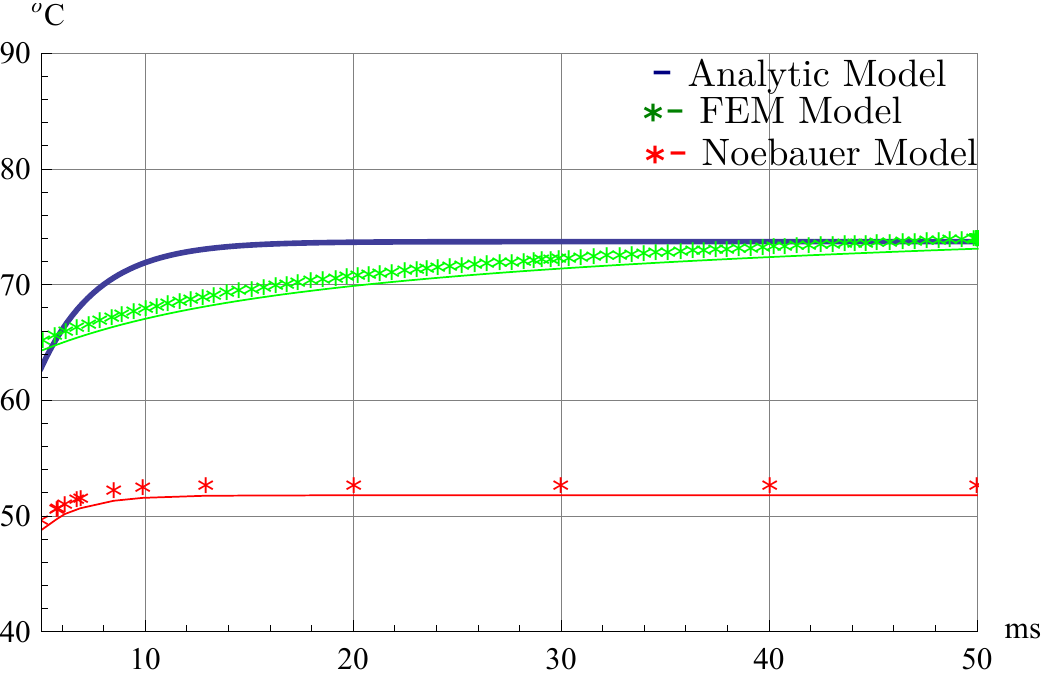}
\caption{Temperature $T_{\text{w}}$ of the wire mid-point 
in time for a gold bondwire of length $L_{\text{w}}=\unit[2.5]{mm}$ and diameter
$D_{\text{w}}=\unit[2.0]{mil}$. The current amplitude amounts to $I_{0}=\unit[2]{A}$.}
\label{bwire_comp}
\end{figure} 

Fig.~\ref{bwire_comp} shows the 
a comparison between our analytic model and
an implementation of Noebauer's model~\cite{bwire_bib3}.
For the reference, we have also included the solution 
obtained via a 3-D FEM simulation of 
the single bondwire heat problem depicted in
Fig.~\ref{bwire_fig1}. Clearly, the model 
in \cite{bwire_bib3} underestimates
the temperature profile. Similar results have been obtained for 
the other wire materials, namely copper and aluminium.
Having verified numerically our solution, in the 
next section we conduct the optimisation
of our model.

\section{Bondwire Model Optimisation}

As in any analytic model built upon 
simplifications, optimisation
based on experimental data is necessary. 
In a real package, where bondwires are 
tightly disposed, the overall
temperature distribution
is governed by a non-linear 
thermo-electromagnetic coupling mechanism.
Furthermore, the manufacturing imperfections, 
mould compound degradation, and current-induced 
ageing of the wires add uncertainties
that are certainly not accounted for by a 
simplified analytic model. 
Hence, to counteract the accumulated 
effect of these uncertainties in the 
accuracy of the model, 
we carry out parameter optimisation in this section.

\subsection{Optimisation Problem Statement}

Let us consider the bondwire model 
as a function 
$\mathrm{B}_{\text{w}}:\underline{\mathbf{p}}\times\left(I_{0},t_{\text{p}}\right)\longmapsto T_{\text{w}}$
with $\underline{\mathbf{p}}=\left[\mathrm{p}_{1},\mathrm{p}_{2},\ldots,\mathrm{p}_{N_{\text{p}}}\right]^{\trans}$;
namely a mapping from a parameter space
$\underline{\mathbf{p}}$ into the temperature
$T_{\text{w}}$ for a given pair
$\left(I_{0},t_{\text{p}}\right)$ of the 
amplitude and duration
of the current. Experimentally, we have collected
a set of data $\left\lbrace I_{0,i},t_{\text{p},i}\right\rbrace$,
with $i=1,\ldots,N_{\text{d}}$ that represents 
fusing events, that is the current amplitude
$I_{0}$ with duration $t_{\text{p}}$ that causes
a bondwire sample to fuse. In this optimisation
we assume that fusing occurs at the wire mid-point
where the hottest spot is expected \cite{bwire_bib4,bwire_bib5,bwire_bib8}. 
In this manner, we define
the model residual
$\mathrm{R}_{\text{w},i}$ as follows
\begin{equation}
\label{bwire_eq33}
\mathrm{R}_{\text{w},i}(\underline{\mathbf{p}})\coloneqq\lvert T_{\text{w;f}}-\mathrm{B}_{\text{w}}\left(\underline{\mathbf{p}},I_{0,i},t_{\text{p},i}\right)\rvert^{2},   
\end{equation} where $T_{\text{w;f}}$
is the corresponding fusing temperature,
and $\mathrm{B}_{\text{w}}\left(\underline{\mathbf{p}},I_{0,i},t_{\text{p},i}\right)$
is the model estimation at the mid-point 
for a given $\left(I_{0,i},t_{\text{p},i}\right)$.
Hence, the total residual for a set of experimental 
data will be
\begin{equation}
\label{bwire_eq34}
\begin{split}
\mathrm{R}_{\text{w}}\left(\underline{\mathbf{p}}\right)&=
\sum_{i=1}^{N_{\text{d}}}\mathrm{R}_{\text{w},i}(\underline{\mathbf{p}})\\
                                                        &=\left(\underline{\mathbf{T}}_{\text{w}}-\underline{\mathbf{B}}_{\text{w}}\left(\underline{\mathbf{p}}\right)\right)^{\trans}
\cdot\left(\underline{\mathbf{T}}_{\text{w}}-\underline{\mathbf{B}}_{\text{w}}\left(\underline{\mathbf{p}}\right)\right)\\
                                                        &=\underline{\mathbf{T}}_{\text{w}}^{\trans}\cdot\underline{\mathbf{T}}_{\text{w}}
-2\underline{\mathbf{T}}_{\text{w}}^{\trans}\cdot\underline{\mathbf{B}}_{\text{w}}\left(\underline{\mathbf{p}}\right) 
+\underline{\mathbf{B}}_{\text{w}}\left(\underline{\mathbf{p}}\right)^{\trans}\cdot
\underline{\mathbf{B}}_{\text{w}}\left(\underline{\mathbf{p}}\right), 
\end{split}
\end{equation} with 
\begin{alignat}{2}
\label{bwire_eq35}
\underline{\mathbf{T}}_{\text{w}}&=
\begin{bmatrix}
T_{\text{w;f}}\\
T_{\text{w;f}}\\
\vdots\\
T_{\text{w;f}}
\end{bmatrix},&\quad
\underline{\mathbf{B}}_{\text{w}}\left(\underline{\mathbf{p}}\right)&=
\begin{bmatrix}
\mathrm{B}_{\text{w}}\left(\underline{\mathbf{p}},I_{0,1},t_{\text{p},1}\right)\\
\mathrm{B}_{\text{w}}\left(\underline{\mathbf{p}},I_{0,2},t_{\text{p},2}\right)\\
\vdots\\
\mathrm{B}_{\text{w}}\left(\underline{\mathbf{p}},I_{0,N_{\text{d}}},t_{\text{p},N_{\text{d}}}\right)
\end{bmatrix}.
\end{alignat}Here, we want to determine 
$\underline{\mathbf{p}}^{\star}$ that minimises 
\eqref{bwire_eq34}, viz.
\begin{equation}
\label{bwire_eq36}
\underline{\mathbf{p}}^{\star}\equiv\argmin{\underline{\mathbf{p}}}
\mathrm{R}_{\text{w}}\left(\underline{\mathbf{p}}\right).   
\end{equation} 

\subsection{Parameter Sub-Set Identification}

We solve for $\underline{\mathbf{p}}^{\star}$ in
\eqref{bwire_eq36} by using 
a Newton-Raphson method \cite{bwire_bib14}
with an order reduction strategy that splits
the space $\underline{\mathbf{p}}$
into well- and ill-conditioned
parameters~\cite{bwire_bib15,bwire_bib16,bwire_bib17}. 
In this manner, robust optimisation 
is accomplished. 

The Newton-Raphson method 
for our multidimensional optimisation problem 
is characterised by the system
\begin{equation}
\label{bwire_eq37}
\underline{\underline{\mathbf{H}}}_{\mathrm{R}_{\text{w}}}\cdot
\Delta\underline{\mathbf{p}}=-\underline{\mathbf{J}}_{\mathrm{R}_{\text{w}}}^{\trans},     
\end{equation} where $\Delta\underline{\mathbf{p}}=\underline{\mathbf{p}}_{k+1}-\underline{\mathbf{p}}_{k}$
is the difference between the \emph{next}
estimation $\underline{\mathbf{p}}_{k+1}$
and the \emph{actual} one $\underline{\mathbf{p}}_{k}$,
$\underline{\mathbf{J}}_{\mathrm{R}_{\text{w}}}$ and 
$\underline{\underline{\mathbf{H}}}_{\mathrm{R}_{\text{w}}}$ are the 
\textit{Jacobian} and \textit{Hessian}
matrices of the residual \cite{bwire_bib18}
evaluated at $\underline{\mathbf{p}}_{k}$.
Let us now assume that we have carried out the 
SVD of $\underline{\underline{\mathbf{H}}}_{\mathrm{R}_{\text{w}}}$
and have ordered the singular values within
$\underline{\underline{\mathbf{\Sigma}}}_{\mathrm{R}_{\text{w}}}$ in
decreasing order of magnitude. Then we nullify the entries 
of the smallest singular values in $\underline{\underline{\mathbf{\Sigma}}}_{\mathrm{R}_{\text{w}}}$  
so as to approximate 
\begin{equation}
\label{bwire_eq39}
\underline{\underline{\mathbf{\Sigma}}}_{\mathrm{R}_{\text{w}}}\approx
\underline{\underline{\widetilde{\mathbf{\Sigma}}}}_{\mathrm{R}_{\text{w}}}=
\begin{bmatrix}
\underline{\underline{\mathbf{\Sigma}}}_{\mathrm{R}_{\text{w}},\text{uu}} &  \underline{\underline{\mathbf{0}}} \\
\underline{\underline{\mathbf{0}}}                               &  \underline{\underline{\mathbf{0}}}          
\end{bmatrix},
\end{equation}that is we cancel out the \emph{diagonal} sub-matrix 
$\underline{\underline{\mathbf{\Sigma}}}_{\mathrm{R}_{\text{w}},\text{ll}}$ 
of \emph{smallest} singular values, and the
order of $\underline{\underline{\widetilde{\mathbf{\Sigma}}}}_{\mathrm{R}_{\text{w}}}$
determines the number of well-conditioned 
parameters. Approximation \eqref{bwire_eq39} leads to
\begin{equation}
\label{bwire_eq40}
\underline{\underline{\mathbf{H}}}_{\mathrm{R}_{\text{w}}}\approx
\underline{\underline{\widetilde{\mathbf{H}}}}_{\mathrm{R}_{\text{w}}}=
\underline{\underline{\mathbf{V}}}_{\mathrm{R}_{\text{w}},\text{1u}}\cdot
\underline{\underline{\mathbf{\Sigma}}}_{\mathrm{R}_{\text{w}},\text{uu}}\cdot    
\underline{\underline{\mathbf{V}}}_{\mathrm{R}_{\text{w}},\text{u1}}^{\text{H}},
\end{equation} where $\underline{\underline{\mathbf{V}}}_{\mathrm{R}_{\text{w}},\text{1u}}$
is the matrix stemming from $\underline{\underline{\mathbf{V}}}_{\mathrm{R}_{\text{w}}}$ 
after cancelling out the singular vectors
associated with the nullified singular values.

To identify the well-conditioned 
parameters of \eqref{bwire_eq37}, 
we perform the QR-decomposition
(with permutation) of
$\underline{\underline{\mathbf{V}}}_{\mathrm{R}_{\text{w}},\text{u1}}^{\text{H}}$ \cite{bwire_bib18}, viz.
\begin{alignat}{2}
\label{bwire_eq41}
\underline{\underline{\mathbf{V}}}_{\mathrm{R}_{\text{w}},\text{u1}}^{\text{H}}\cdot
\underline{\underline{\mathbf{P}}}&=\underbrace{ 
\begin{bmatrix}
\underline{\underline{\mathbf{V}}}_{\mathrm{R}_{\text{w}},\text{uu}}^{\text{H}} &
\underline{\underline{\mathbf{V}}}_{\mathrm{R}_{\text{w}},\text{ul}}^{\text{H}}
\end{bmatrix}}_{\underline{\underline{Q}}\cdot\underline{\underline{R}}}\Rightarrow &\;
\underline{\underline{\mathbf{P}}}^{\trans}\cdot   
\underline{\underline{\mathbf{V}}}_{\mathrm{R}_{\text{w}},\text{1u}}&=
\underbrace{\begin{bmatrix}
\underline{\underline{\mathbf{V}}}_{\mathrm{R}_{\text{w}},\text{uu}}\\
\underline{\underline{\mathbf{V}}}_{\mathrm{R}_{\text{w}},\text{lu}}
\end{bmatrix}}_{\underline{\underline{R}}^{\trans}\cdot\underline{\underline{Q}}^{\trans}}.
\end{alignat} Above, the \emph{permutation}
matrix $\underline{\underline{\mathbf{P}}}$ swaps columns of 
$\underline{\underline{\mathbf{V}}}_{\mathrm{R}_{\text{w}},\text{u1}}^{\text{H}}$
to split them into a set of linearly
\emph{independent} singular vectors
$\underline{\underline{\mathbf{V}}}_{\mathrm{R}_{\text{w}},\text{uu}}^{\text{H}}$ 
and a set of linearly \emph{dependent}
singular vectors $\underline{\underline{\mathbf{V}}}_{\mathrm{R}_{\text{w}},\text{ul}}^{\text{H}}$.
This splitting permits to separate the set of well-posed parameters
from the ill-posed ones. Identities \eqref{bwire_eq40}
and \eqref{bwire_eq41} lead to the sought
reduced counterpart of \eqref{bwire_eq37}, viz.
\begin{equation}
\label{bwire_eq42}
\underline{\underline{\widetilde{\mathbf{H}}}}_{\mathrm{R}_{\text{w}},\text{uu}}\cdot
\Delta\underline{\widetilde{\mathbf{p}}}_{\text{u}}=-\underline{\mathbf{J}}_{\mathrm{R}_{\text{w}},\text{u}}^{\trans},     
\end{equation} with
\begin{alignat}{2}
\underline{\underline{\widetilde{\mathbf{H}}}}_{\mathrm{R}_{\text{w}},\text{uu}}&=
\underline{\underline{\mathbf{V}}}_{\mathrm{R}_{\text{w}},\text{uu}}\cdot
\underline{\underline{\mathbf{\Sigma}}}_{\mathrm{R}_{\text{w}},\text{uu}}\cdot    
\underline{\underline{\mathbf{V}}}_{\mathrm{R}_{\text{w}},\text{uu}}^{\text{H}};&\quad
\underline{\underline{\mathbf{P}}}^{\trans}\cdot   
\Delta\underline{\mathbf{p}}&=
\begin{bmatrix}
\Delta\underline{\widetilde{\mathbf{p}}}_{\text{u}}\\
\underline{\mathbf{0}}
\end{bmatrix};\nonumber
\end{alignat}
\begin{equation}
\label{bwire_eq43}
\underline{\underline{\mathbf{P}}}^{\trans}\cdot   
\underline{\mathbf{J}}_{\mathrm{R}_{\text{w}}}^{\trans}=
\begin{bmatrix}
\underline{\mathbf{J}}_{\mathrm{R}_{\text{w}},\text{u}}^{\trans}\\
\underline{\mathbf{J}}_{\mathrm{R}_{\text{w}},\text{l}}^{\trans}
\end{bmatrix},
\end{equation} where we group and
retain in \eqref{bwire_eq42} the entries corresponding 
to the well-posed parameters, while setting 
constant the ill-posed ones by stating that  
$\Delta\underline{\widetilde{\mathbf{p}}}_{\text{l}}=\underline{\mathbf{0}}$
in \eqref{bwire_eq43}.

Finaly, we express
$\underline{\mathbf{J}}_{\mathrm{R}_{\text{w}}}$ and 
$\underline{\underline{\mathbf{H}}}_{\mathrm{R}_{\text{w}}}$ 
in terms of $\underline{\underline{\mathbf{J}}}_{\mathrm{B}_{\text{w}}}$   
and $\underline{\underline{\mathbf{H}}}_{\mathrm{B}_{\text{w}}}$; namely, 
the Jacobian and Hessian matrices of the 
bondwire model $\mathrm{B}_{\text{w}}$. 
Further mathematical analysis shows that these matrices 
are related as follows
\begin{align}
\label{bwire_eq44}
\underline{\mathbf{J}}_{\mathrm{R}_{\text{w}}}&=2\left(\underline{\mathbf{B}}^{\trans}_{\text{w}}\left(\underline{\mathbf{p}}\right)
-\underline{\mathbf{T}}^{\trans}_{\text{w}}\right)\cdot
\underline{\underline{\mathbf{J}}}_{\mathrm{B}_{\text{w}}},\\
\label{bwire_eq45}
\underline{\underline{\mathbf{H}}}_{\mathrm{R}_{\text{w}}}&=2\underline{\underline{\mathbf{J}}}^{\trans}_{\mathrm{B}_{\text{w}}}\cdot
\underline{\underline{\mathbf{J}}}_{\mathrm{B}_{\text{w}}}+2\sum_{i=1}^{N_{\text{d}}}
\left(\mathrm{B}_{\text{w}}\left(\underline{\mathbf{p}},I_{0,i},t_{\text{p},i}\right)-T_{\text{w;f}}\right)
\underline{\underline{\mathbf{H}}}_{\mathrm{B}_{\text{w}},i}.
\end{align}

In the next section, we briefly describe the
experimental setup used to generate 
and measure the required bondwire fusing events.

\subsection{Experimental Setup}

\begin{figure}[htbp!]
\centering
\includegraphics[width=0.75\columnwidth]{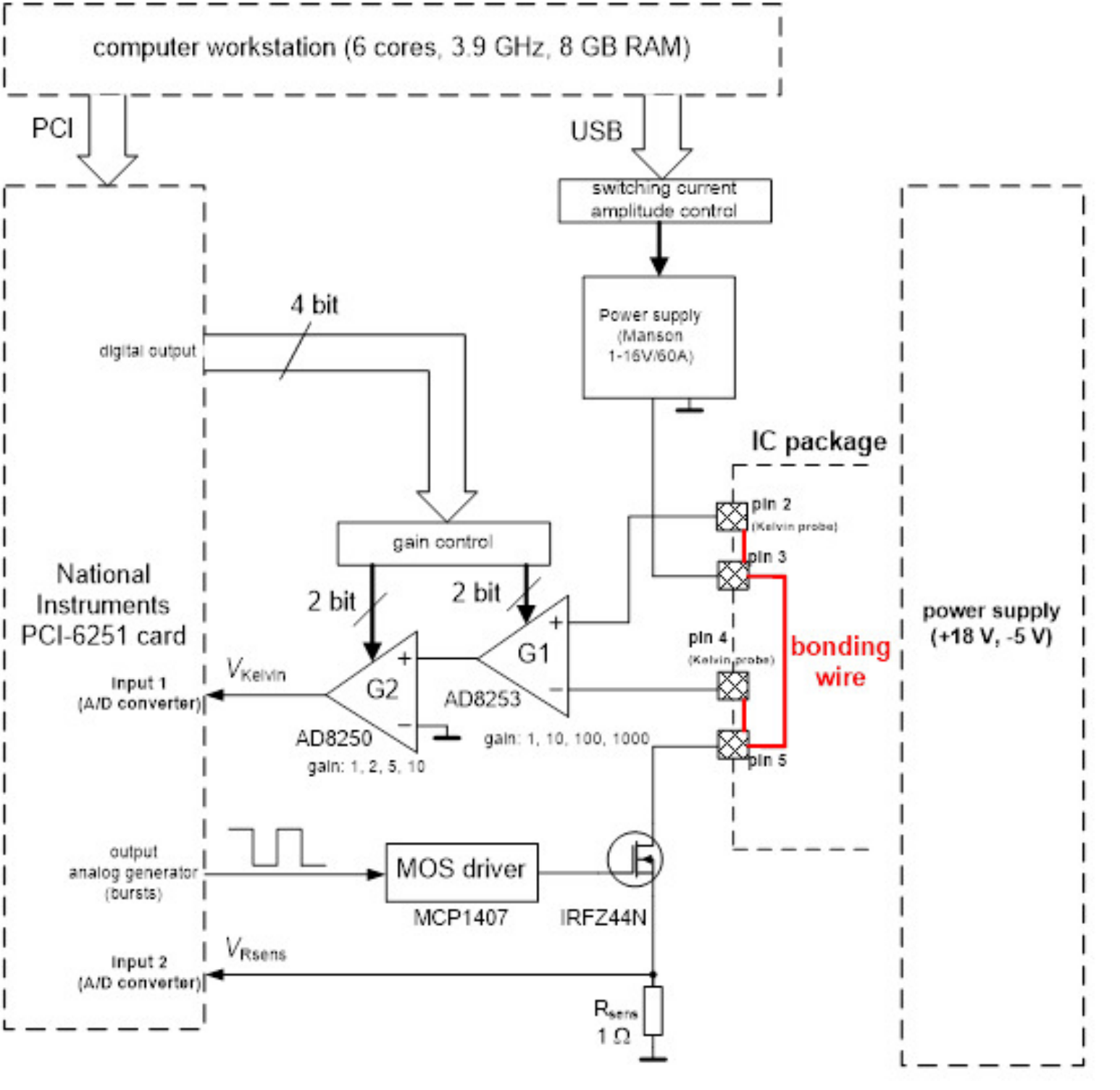}
\caption{Schematic of the bondwire tester.} 
\label{bwire_fig7}
\end{figure}

The experimental setup for the extraction of
fusing events and the dynamical parameters
of the encapsulated bondwires
is described in this section.
In Fig.~\ref{bwire_fig7} we show the 
schematic of the so-called bondwire tester. 
The setup permits statistical quantification of  
the current capabilities and fatigue 
of the bondwires. All integrated circuits and
Kelvin probes were fabricated by ON-Semiconductor (Belgium). 

The tester consists of a
power supply ($\unit[16]{V}$ and $\unit[60]{A}$), a test interface
(custom made board), a personal computer
(with MATLAB\texttrademark), and a data acquisition card (DAQ)
(National Instruments 6251) capable
of capturing $\unit[1]{MS/s}$ multichannel, $\unit[16]{bit}$ resolution
and with a voltage range of $\pm\unit[10]{V}$.
The test interface consists
of six power channels (for six bondwire pairs). 
To avoid using 12 acquisition channels, each channel 
is addressed by demultiplexing the driving signal to a specific bondwire.
Thus, the current flowing through 
and the voltage drop along the bondwire
are sensed, amplified, and multiplexed 
to outputs which are digitised by a NI
card and thus returned back to the computer.
\begin{figure}[htbp!]
\centering
\includegraphics[width=0.65\columnwidth]{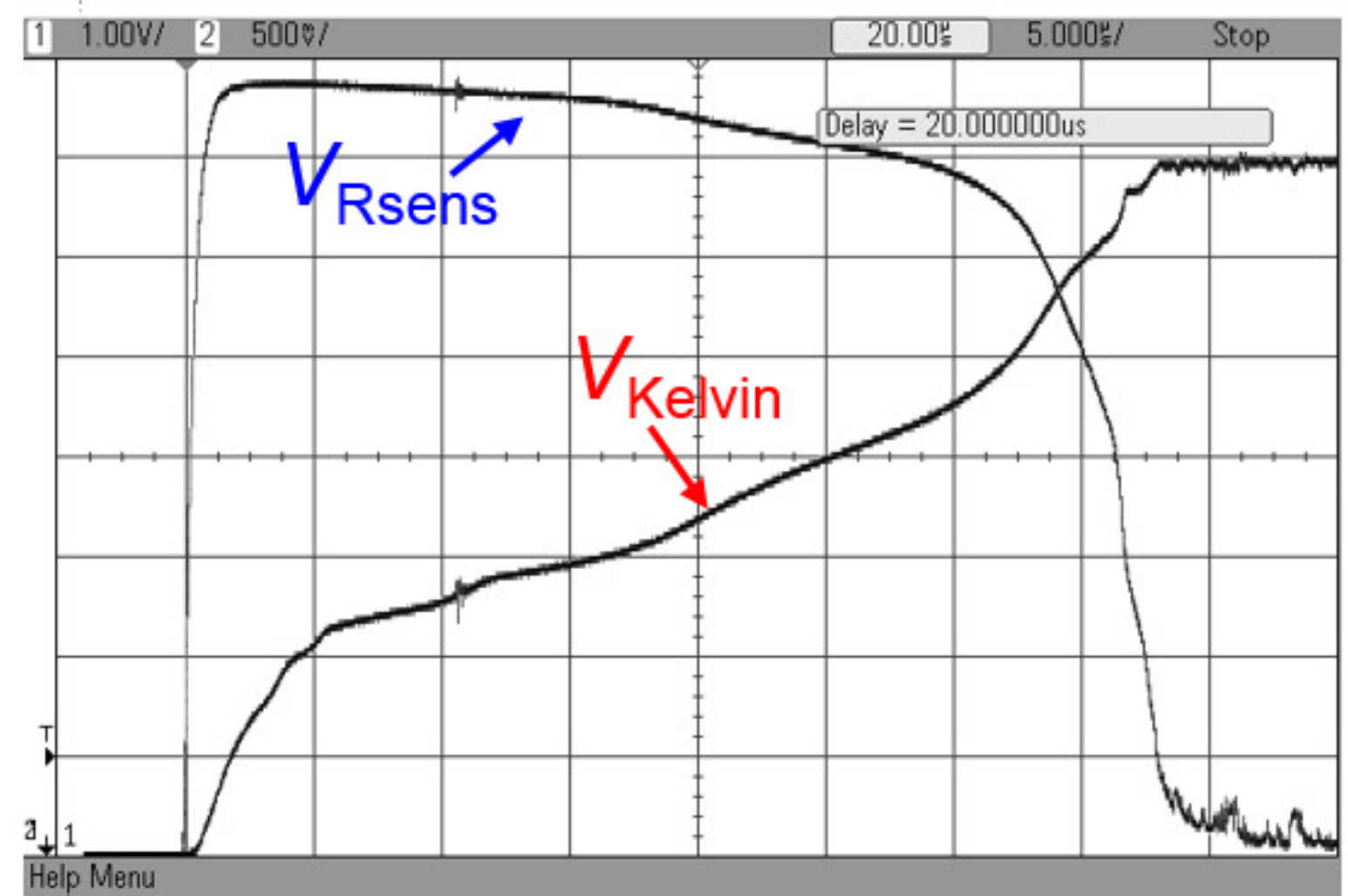}
\caption{Details of $V_{\text{Rsens}}$ and $V_{\text{Kelvin}}$
captured with the oscilloscope.} 
\label{bwire_fig8}
\end{figure}

In general, the experimental setup can be divided into
two parts, namely the software which directly controls the NI
card and consequently the tester, and the printed circuit board (PCB)
of the tester with the 
multichannel switching power source. 
Each bondwire is addressed by the
analog multiplexer (HCF4051BE) upon selecting
the required channel among six possibilities.
The PCB contains six power
switches (MOSFET transistors IRFZ044); the same
amount of amplifiers are connected
to the bondwires via Kelvin probes. The
amplifiers have differential voltage inputs and non-symmetric
outputs with a digitally controlled gain (see Fig.~\ref{bwire_fig7}).
Since a high final gain is required, 
a two-stage cascade connection is employed. 
Up to four bits, which are directly set via a MATLAB script,
can be used to set the final gain factor. 
\begin{figure}[htbp!]
\centering
\includegraphics[width=0.65\columnwidth]{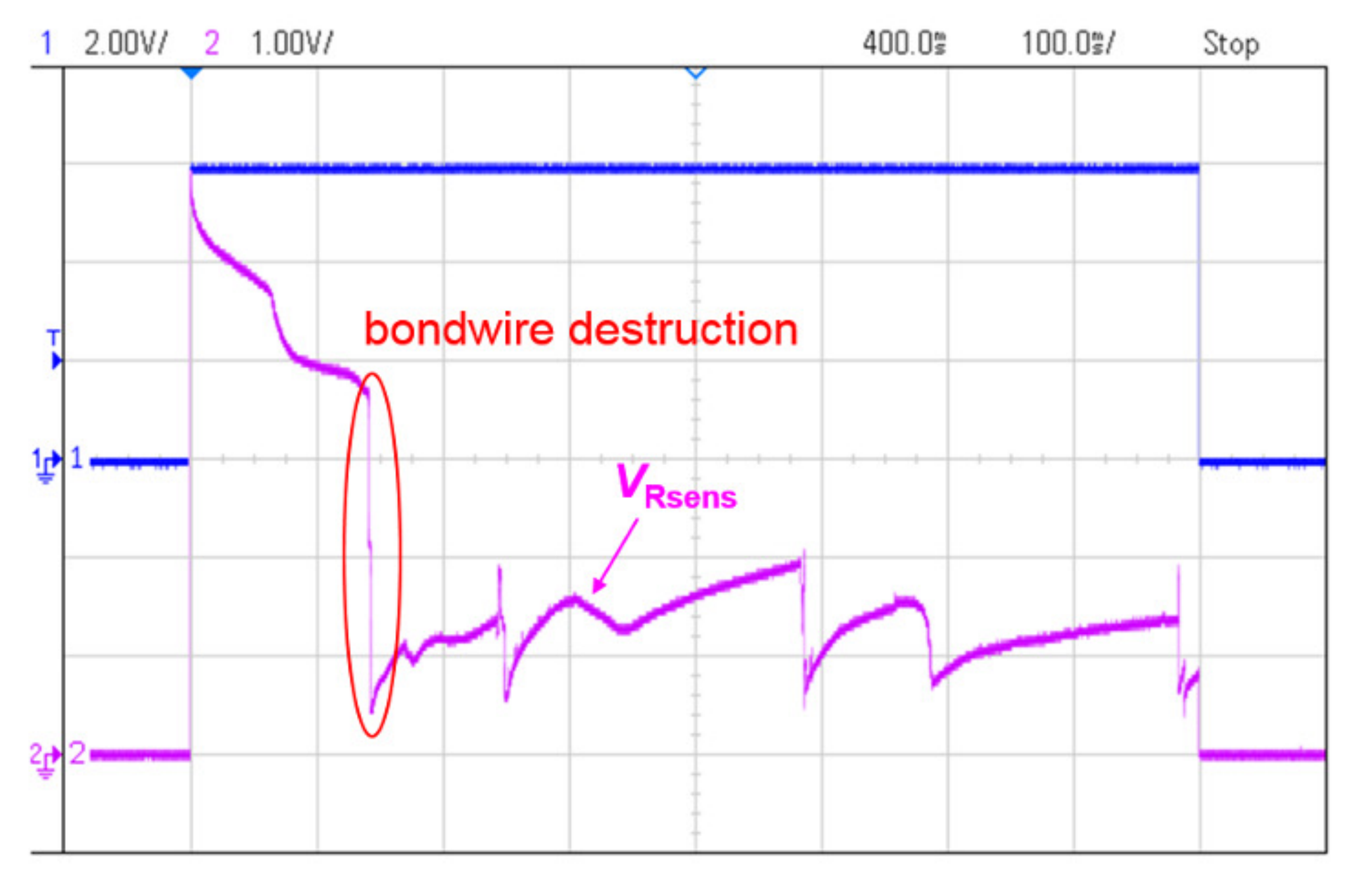}
\caption{Dynamic behaviour after fusing of the bondwire.} 
\label{bwire_fig9}
\end{figure}

To amplify the
voltage $V_{\text{Kelvin}}$
sensed by the probes (see Fig.~\ref{bwire_fig7}),
a high-gain stage (AD8253 PGA) with gain $\in\,\lbrace 1, 10, 100, 1000\rbrace$
is combined with an integrated circuit (AD8250) whose gain $\in\,\lbrace1, 2, 5, 10\rbrace$.
All power transistors have their sources connected to a single sensing resistance of 
$\unit[1]{\Omega}\,(\unit[50]{W})$ that measures the
voltage drop $V_{\text{Rsens}}$ inasmuch as 
the dominant current will always come from the 
branch of the active
bondwire, while contributions from the other wires are negligible.
Additionally a ringing snubber is added to 
compensate overshoots created by switching 
of an inductive load. The time evolution of 
$V_{\text{Kelvin}}$ and $V_{\text{Rsens}}$
is monitored with the oscilloscope (see Fig.~\ref{bwire_fig8}).
In parallel to the outputs $V_{\text{Rsens}}$ and
$V_{\text{Kelvin}}$, Zener diodes
(voltage limiters) are used to avoid 
damage of the external A/D 
converter when used in a sensitive range.

In Fig.~\ref{bwire_fig9} we show a selected 
screenshot of a fusing event that does not translate immediately
into an open circuit. Various degradation processes 
have been identified with the delayed 
time-base features of the oscilloscope. 
\begin{figure}[!htbp]
\centering
\includegraphics[width=0.70\columnwidth]{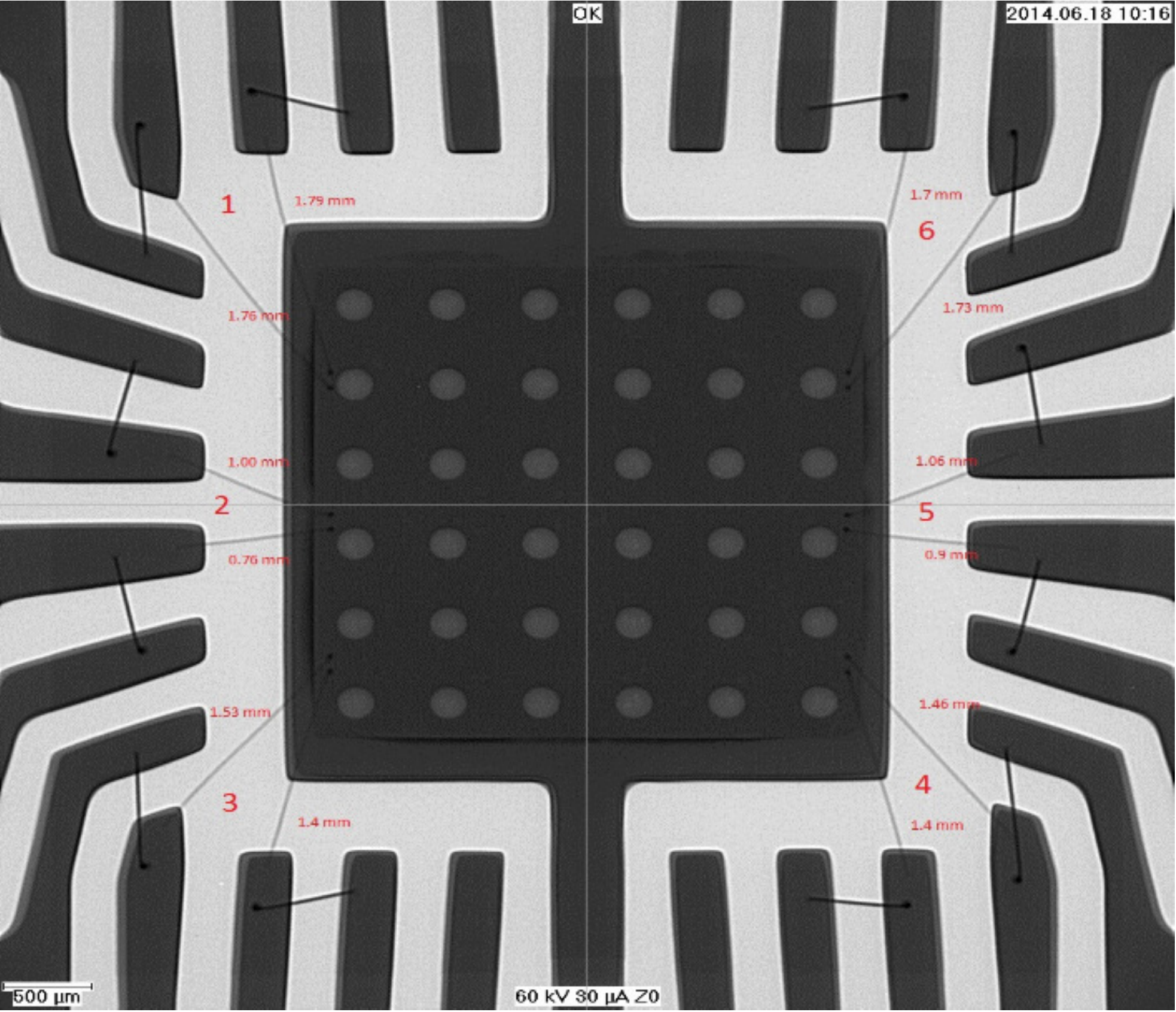}
\caption{Example of an X-ray picture of a single package.} 
\label{bwire_fig10}
\end{figure}

In Fig.~\ref{bwire_fig10} we show an X-ray picture of a
sample package. Each sample package consists of 6 pairs of
bondwires as enumerated in the picture. The length 
of each (in $\unit[]{mm}$) is also indicated therein.
A fusing event is induced by using a long duration pulse.
The characteristics of the pulse are 
controlled by the power 
supply Manson (capable of delivering up to $\unit[60]{A}$ at
$\unit[16]{V}$) in order to obtain
certain amount of current flowing through the bondwire.
The time required to fuse the bondwire
is measured
and the results are then provided  
for all available types, that is material, diameter, and 
position within the package.

\subsection{Optimisation Results}

\begin{figure}[htbp!]
\centering
\subfloat[]{\label{bwire_fig6a}
\includegraphics[width=0.540\columnwidth]{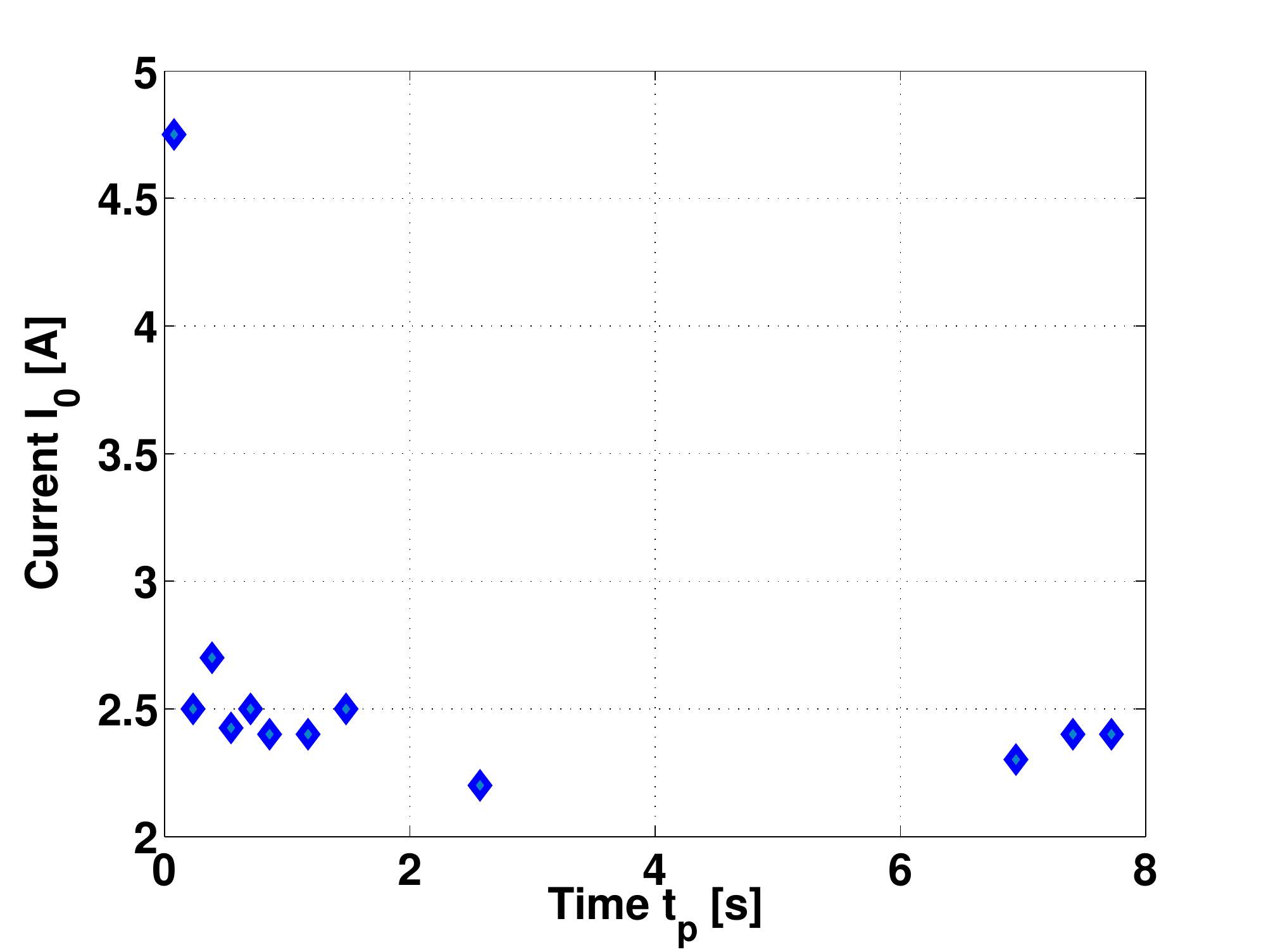}}
\subfloat[]{\label{bwire_fig6b}
\includegraphics[width=0.540\columnwidth]{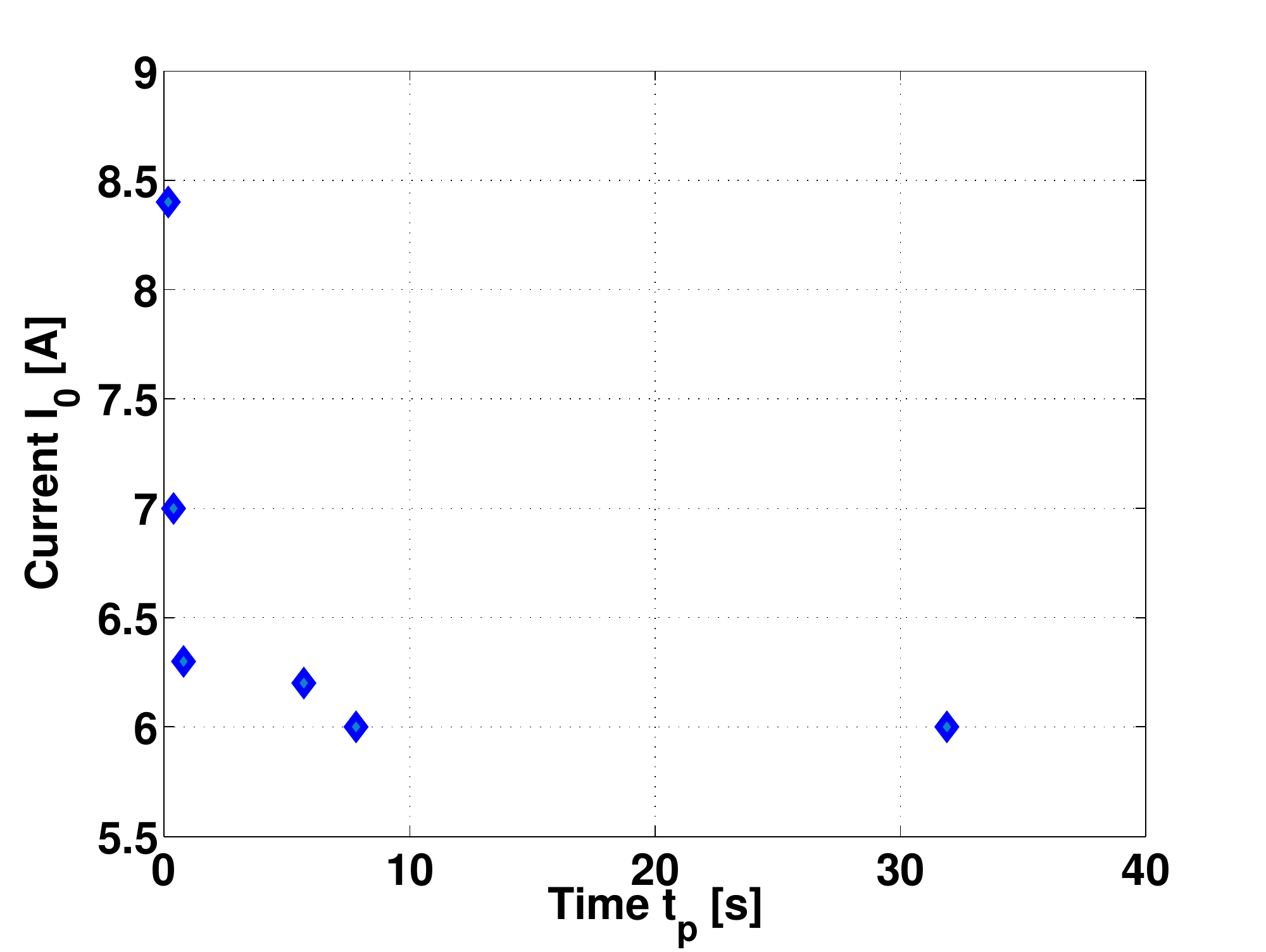}}\\
\caption{\label{bwire_fig6} Experimental 
filtered data 
$\lbrace\bar{I}_{0,i},\bar{t}_{\text{p},i}\rbrace$ 
for gold and copper bondwires at position 1 
in a package.~(a) $\mathrm{Au}$-wire of
$L_{\text{w}}=\unit[1.712]{mm}$ and 
$D_{\text{w}}=\unit[1.0]{mil}$;~(b) $\mathrm{Cu}$-wire
of $L_{\text{w}}=\unit[3.450]{mm}$
and $D_{\text{w}}=\unit[2.0]{mil}$.}
\end{figure}

In this section we apply the afore-described  
parameter sub-set identification method to
carry out the optimisation of the bondwire model.
As stated before, each bondwire sample within a
package (see Fig.~\ref{bwire_fig10}) is characterised  
by a set of data points representing fusing events. 
Each fusing event corresponds to a pair
$\lbrace I_{0,i},t_{\text{p},i}\rbrace$    
representing the current amplitude $I_{0,i}$
that for given time $t_{\text{p},i}$ fuses
the wire. In total 1077 bondwire samples within
their package were deliberately fused in the data 
collection phase. Each data set is then filtered 
by means of a histogram to generate 
a sequence of smoother
pairs $\lbrace\bar{I}_{0,i},\bar{t}_{\text{p},i}\rbrace$ 
that are used in the optimisation.
Fig.~\ref{bwire_fig6} shows two examples 
of the resulting filtered data
$\lbrace\bar{I}_{0,i},\bar{t}_{\text{p},i}\rbrace$ 
for $\mathrm{Au}$- and $\mathrm{Cu}$-wires of
diameters $\unit[1.0]{mil}$ and 
$\unit[2.0]{mil}$ at position 1 in the package
(see Fig.~\ref{bwire_fig10}).

 Similar data is obtained
for the bondwires at the other positions within
the package. In Table.~\ref{bwire_tab1}
and \ref{bwire_tab2}, we 
collect the nominal geometric, electric, and 
thermic values of the bondwires. The nominal 
diameter $\bar{D}_{\text{w}}$ were provided 
by the package manufacturer, while 
the nominal length $\bar{L}_{\text{w}}$
has been obtained through averaging
of the length measured via the x-ray
scanning (see Fig.~\ref{bwire_fig10})     
and the length inferred from the lead-frame data-sheet.

\begin{table}[htb!]
\caption{Geometric characteristics of
the bondwires within a package. The nominal $\bar{D}_{\text{w}}$
and $\bar{L}_{\text{w}}$ are given in $\unit[]{[mil]}$ and 
$\unit[]{[mm]}$.}
\label{bwire_tab1}
\centering
\begin{tabular}{c|c|c|c|c|c}
\hline
\multirow{2}{*}{}&\multicolumn{3}{c|}{$\mathbf{Cu}$\textbf{-wire}} & $\mathbf{Au}$\textbf{-wire} & \multirow{2}{*}{\textbf{Pos}}\\
\cline{2-5}
          & $\unit[1.0]{mil}$ & $\unit[1.3]{mil}$  & $\unit[2.0]{mil}$  & $\unit[1.0]{mil}$ &    \\
\hline
\multirow{6}{*}{$\bar{L}_{\text{w}}$ ~$\unit[]{[mm]}$} & $\unit[2.025]{}$ & $\unit[3.445]{}$ & $\unit[3.450]{}$ & $\unit[1.712]{}$& $1$ \\  
       & $\unit[1.267]{}$ & $\unit[2.160]{}$ & $\unit[2.160]{}$ & $\unit[1.226]{}$& $2$ \\  
       & $\unit[1.696]{}$ & $\unit[3.458]{}$ & $\unit[3.458]{}$ & $\unit[1.889]{}$& $3$ \\  
       & $\unit[1.670]{}$ & $\unit[3.469]{}$ & $\unit[3.469]{}$ & $\unit[1.938]{}$& $4$ \\
       & $\unit[1.114]{}$ & $\unit[2.126]{}$ & $\unit[2.126]{}$ & $\unit[1.369]{}$& $5$ \\
       & $\unit[1.884]{}$ & $\unit[3.434]{}$ & $\unit[3.434]{}$ & $\unit[1.827]{}$& $6$ \\
\hline
\end{tabular}
\end{table}
\begin{table}[htb!]
\caption{Nominal electric and thermic characteristics
of the bondwires.}
\label{bwire_tab2}
\centering
\begin{tabular}{c|c|c}
\hline
\textbf{Elec} & $\mathbf{Cu}$\textbf{-wire} & $\mathbf{Au}$\textbf{-wire}\\  
\hline
$\bar{\uprho}_{\text{e;o}}\unit[]{[\Omega.m]}$ & $1.678.10^{-8}$ & $2.214.10^{-8}$ \\ 
\hline
$\bar{\alpha}_{\uprho}\unit[]{[\frac{1}{K}]}$ & $3.862.10^{-3}$ & $3.400.10^{-3}$ \\
\hline
\textbf{Therm} & \multicolumn{2}{c}{} \\
\hline
$\bar{\rho}_{\text{w}}\unit[]{[\frac{Kg}{m^{3}}]}$ & $8960$ & $19300$ \\
\hline
$\bar{\kappa}_{\text{o}}\unit[]{[\frac{W}{(m.K)}]}$ & $398$ & $315$ \\ 
\hline
$\bar{\alpha}_{\kappa}\unit[]{[\frac{1}{K}]}$ & $-4.675.10^{-4}$ & $-2.744.10^{-4}$ \\
\hline
$\bar{c}_{\text{e;w}}\unit[]{[\frac{J}{Kg.K}]}$ & $353$ & $129$ \\
\hline
$\bar{\epsilon}_{\text{w}}$ & $3.750.10^{-2}$ & $2.475.10^{-1}$ \\
\hline
$\bar{T}_{\text{ch}}\unit[]{[K]}$ & $300.5$ & $300.5$ \\
\hline
$\bar{T}_{\text{ld}}\unit[]{[K]}$ & $300.5$ & $300.5$ \\
\hline
\end{tabular}
\end{table} The electric and thermic values collected
in Table~\ref{bwire_tab2} are standard for these
materials~\cite{bwire_bib19}. In particular,
the value adopted for the emissivity 
$\bar{\epsilon}_{\text{w}}$ corresponds to 
polished copper and gold wires; nevertheless, for
bondwires that are embedded in real packages, this polishness
might not be the case. These values,
together with the filtered data $\lbrace\bar{I}_{0,i},\bar{t}_{\text{p},i}\rbrace$, 
are substituted in \eqref{bwire_eq44}--\eqref{bwire_eq45} 
whence the well-posed parameters are established as explained 
above. In carrying out the optimisation of these parameters, 
we have adhered to the following principles stemming
from the physicality of the problem.
\begin{itemize}
\item The bondwire length $L_{\text{w}}$ and
diameter $D_{\text{w}}$ are always allowed 
to vary up to $30\%$ around their nominal values.
This relative variation is the add up of the 
manufacturing tolerance, the tolerance in the 
x-ray scanning, the tolerance in the data-sheet,
and a margin of safety.
\item The chip and lead temperatures $T_{\text{ch}}$
and $T_{\text{ld}}$, although fixed as isothermal 
boundary conditions within the model, are allowed
to vary up to $50\%$ above their nominal values.
This is done to accommodate
for the actual temperature increase 
these two bodies may undergo during
a fusing event. This increase of 
temperature is certainly not negligible       
for fusing events characterised by
long times to fuse.
\item The dimensions of the moulding compound
section around a bondwire at a given position,
as required by the model, are directly taken
from the package dimensions.
\item The convective coefficient
$h_{\text{c}}$ is assumed to be for air at 
normal conditions given that no special 
fluid dynamic condition in the room 
was expected. 
\item The remaining parameters are allowed to 
vary freely in a interval that makes physical sense, e.g.,
the emissivity is such that
$0<\epsilon_{\text{w}}\leq1$, while $\alpha_{\uprho}\geq0$
and $\alpha_{\kappa}\leq0$.
\end{itemize}

\begin{table}[htb!]
\caption{Overall average variation after optimisation 
of the parameters defining the bondwire model.}
\label{bwire_tab3}
\centering
\begin{tabular}{c|c|c|c|c|c}
\hline
\multirow{2}{*}{\textbf{Geom}} &\multicolumn{3}{c|}{$\mathbf{Cu}$\textbf{-wire}} & $\mathbf{Au}$\textbf{-wire} & \multirow{2}{*}{$\mathbf{\Delta}_{\text{tot}}$} \\
\cline{2-5}
           & $\unit[1.0]{mil}$ & $\unit[1.3]{mil}$ & $\unit[2.0]{mil}$ & $\unit[1.0]{mil}$ &     \\
\hline           
$\Delta\bar{D}_{\text{w}}$ & $-17.53\%$ & $5.88\%$ & $-20.39\%$ & $-26.98\%$ & $\mathbf{-14.76}\%$ \\
\hline
$\Delta\bar{L}_{\text{w}}$ & $12.88\%$ & $-16.06\%$ & $20.33\%$ & $-3.92\%$ & $\mathbf{3.31}\%$ \\  
\hline
\textbf{Elec} & \multicolumn{4}{c}{} \\
\hline
$\Delta\bar{\uprho}_{\text{e;o}}$ & $106.63\%$ & $7.22\%$ & $2.95\%$ & $-18.54\%$ & $\mathbf{24.57}\%$ \\
\hline
$\Delta\bar{\alpha}_{\uprho}$ & $-48.17\%$ & $1.05\%$ & $-21.07\%$ & $-18.75\%$ & $\mathbf{-21.74}\%$ \\
\hline
\textbf{Therm} & \multicolumn{4}{c}{} \\
\hline
$\Delta\bar{\rho}_{\text{w}}$ & $144.43\%$ & $142.44\%$ & $341.53\%$ & $181.43\%$ & $\mathbf{202.46}\%$ \\
\hline
$\Delta\bar{\kappa}_{\text{o}}$ & $16.91\%$ & $0.00\%$ & $-15.00\%$ & $-20.83\%$  & $\mathbf{-4.73}\%$ \\ 
\hline
$\Delta\bar{\alpha}_{\kappa}$ & $36.42\%$ & $-22.64\%$ & $-322.67\%$ & $-23.85\%$   & $\mathbf{-83.19}\%$ \\
\hline
$\Delta\bar{c}_{\text{e;w}}$ & $126.04\%$ & $298.23\%$ & $125.96\%$ & $187.63\%$ & $\mathbf{184.47}\%$ \\
\hline
$\Delta\bar{\epsilon}_{\text{w}}$ & $390.81\%$ & $-32.48\%$ & $168.22\%$ & $12.99\%$ & $\mathbf{134.89}\%$ \\
\hline
$\Delta\bar{T}_{\text{ch}}$ & $19.73\%$ & $10.00\%$ & $24.17\%$ & $30.00\%$ & $\mathbf{20.98}\%$ \\
\hline
$\Delta\bar{T}_{\text{ld}}$ & $30.82\%$ & $18.33\%$ & $34.17\%$ & $35.83\%$ & $\mathbf{29.79}\%$ \\
\hline
\end{tabular}
\end{table}

\begin{table}[htb!]
\caption{Nominal electric and thermic characteristics
of the bondwires within the package.}
\label{bwire_tab4}
\centering
\begin{tabular}{c|c|c|c|c|c}
\hline
\multirow{2}{*}{} &\multicolumn{3}{c|}{$\mathbf{Cu}$\textbf{-wire}} & $\mathbf{Au}$\textbf{-wire} & \multirow{2}{*}{$\pmb{\upvarepsilon}_{\text{tot}}$} \\
\cline{2-5}
           & $\unit[1.0]{mil}$ & $\unit[1.3]{mil}$ & $\unit[2.0]{mil}$ & $\unit[1.0]{mil}$ &     \\
\hline           
$\upvarepsilon^{\text{t}}_{\mathrm{B_{\text{w}}}}$ &  $31.52\%$ & $21.70\%$ & $27.83\%$ & $35.87\%$ & $\mathbf{29.23}\%$ \\
\hline
$\upvarepsilon^{\text{s}}_{\mathrm{B_{\text{w}}}}$ &  $19.78\%$ & $24.31\%$ & $12.88\%$ & $24.22\%$ & $\mathbf{20.30}\%$ \\
\hline
        &\multicolumn{4}{c|}{}&       \\
\hline
$\upvarepsilon^{\star\text{t}}_{\mathrm{B_{\text{w}}}}$ & $6.95\%$ & $3.26\%$ & $3.26\%$ & $8.81\%$ & $\mathbf{5.57}\%$ \\
\hline
$\upvarepsilon^{\star\text{s}}_{\mathrm{B_{\text{w}}}}$ & $3.90\%$ & $4.88\%$ & $2.14\%$ & $11.16\%$& $\mathbf{5.52}\%$ \\
\hline
\end{tabular}
\end{table}

In Table \ref{bwire_tab3} we have collected 
the average relative 
variation, with respect to their nominal values, of
the parameters defining the 
bondwire model after 
optimisation. The variations are the 
average for each bondwire regardless of
its position within the package.
The overall total variation is given in the  
last column of the table. As we may
see, all parameters undergo variations  
because the set of well-posed parameters, i.e.,
the parameters to be optimised, can be different
from one data set to the other. Consequently, 
the parameters that are optimised for
a bondwire at a given 
position may become the ill-posed ones
for the bondwire at the next position.

As to the actual values of the variations,
we first notice that the geometric parameters
of the wire; namely its diameter and length,
do not suffer major changes with respect to their 
nominal values. This implies that the model 
retains the physicality of the problem in this respect. 
Next, we start by 
recalling that the bondwire temperature formula
in \eqref{bwire_eq18} consist of 
two distinguishable components; the first
$\tilde{\theta}_{\text{w};2}(y)$  
that accounts for the \emph{steady} 
regime of the temperature, and second 
$\tilde{\theta}_{\text{w};1}(y,t)$ that
gives account of the \emph{transient}.
By inspection of the corresponding formulas,
we may identify which parameters play predominant
roles in either regime. For instance,
the parameters $\lbrace\rho_{\text{w}},c_{\text{e;w}}\rbrace$
have only influence on the transient (cf.~\eqref{bwire_eq7}),
whereas $\lbrace\kappa_{\text{o}},\alpha_{\kappa},
\uprho_{\text{e};\text{w}},\alpha_{\uprho},T_{\text{ch}},T_{\text{ld}}\rbrace$
have mostly influence on the steady regime. 
Owing to the simplified configuration
of the model and the dynamics of these 
two regimes in a real package, we may expect the model 
to yield better temperature estimations
in the steady regime than in the transient. 
This fact manifest in the relatively high
variation of $\lbrace\rho_{\text{w}},c_{\text{e;w}}\rbrace$ with
respect to their nominal values as compared with
the variation of $\lbrace\kappa_{\text{o}},\alpha_{\kappa},
\uprho_{\text{e};\text{w}},\alpha_{\uprho},T_{\text{ch}},T_{\text{ld}}\rbrace$.
With regard to the wire emissivity $\epsilon_{\text{w}}$,
the overall high relative increase confirms 
that the assumption of polished wires is not 
a good initial guess.

To report on the accuracy of the model
before and after optimisation, in 
Table \ref{bwire_tab4} we collect the
average relative error of the model. The error
therein is the average of the normalised 
residual (cf.~\eqref{bwire_eq34}) for all 
bondwires regardless of their 
position within the package.
We have split the error into the component
$\upvarepsilon^{\text{t}}_{\mathrm{B_{\text{w}}}}$
that quantifies the error in the transient, 
and the component $\upvarepsilon^{\text{s}}_{\mathrm{B_{\text{w}}}}$
that does so in the steady regime. This splitting 
can be done by noticing in \eqref{bwire_eq18}
that any fusing event occurring   
at a time less or equal than the relevant time 
constant is likely in the transient.
The results in Table~\ref{bwire_tab4} confirms 
what has been already inferred from Table~\ref{bwire_tab3}, 
that is the model yields better estimations
for long time excitations. In general,
the performance of the model before optimisation 
is pretty good, and is greatly improved 
with the optimisation.

\section{Conclusion}

We have presented an extended analytic formula
for the estimation of the heating of bondwires
within real packages. The model retains the  
the shape and dimensions of the moulding compound 
surrounding the wire and imposes suitable
boundary conditions for the temperature distribution.
To yield the temperature along the wire, the model couples 
the heat transfer equations of both the wire and 
the compound by means of effective transfer coefficients
that stems from the linearisation of the thermal radiation  
term on the wire surface. The resulting wire 
temperature formula consists of simple  
basic functions that make it suitable for a fast
implementation into a numerical calculator.
The model has also been optimised with a series 
of experimental measurements representing fusing events.
The idea upon which the formula is built can be certainly 
extended to handle several bondwires in a package. 
This would entail defining as many  
effective transfer coefficients as bondwires.

\appendices
\section{Moulding Compound Temperature Functions and Heat Kernel}
\label{bwire_appA}
Here, the functions
comprising the compound temperature expression of 
\eqref{bwire_eq20} are derived. The component
$\widetilde{T}_{\text{m}}\left(x,y,z,t\right)=
\widetilde{T}_{\text{m};1}\left(x,y,z,t\right)+
\widetilde{T}_{\text{m};2;1}\left(x,y,z\right)+\widetilde{T}_{\text{m};2;2}\left(x,y,z\right)$
must satisfy
\begin{equation}
\label{bwire_eq21}
\rho_{\text{m}} c_{\text{e};\text{m}}
\frac{\partial \widetilde{T}_{\text{m}}}{\partial t}=
\kappa_{\text{m}}\nabla^{2}\widetilde{T}_{\text{m}}.
\end{equation} Owing to their definition and the 
$yz$-plane of symmetry in Fig.~\ref{bwire_fig2},
these components are subject to the following 
BCs
\begin{gather}
-\kappa_{\text{m}}\frac{\partial}{\partial z}
\widetilde{T}_{\text{m};1}\left(x,y,\frac{H_{\text{m}}}{2},t\right)= 
h_{\text{c}}\widetilde{T}_{\text{m};1}\left(x,y,\frac{H_{\text{m}}}{2},t\right);\nonumber\\ 
\label{bwire_eq22}
-\kappa_{\text{m}}\frac{\partial}{\partial z}
\widetilde{T}_{\text{m};2,1}\left(x,y,\frac{H_{\text{m}}}{2}\right)= 
h_{\text{c}}\widetilde{T}_{\text{m};2,1}\left(x,y,\frac{H_{\text{m}}}{2}\right);\\
-\kappa_{\text{m}}\frac{\partial}{\partial z}
\widetilde{T}_{\text{m};2,2}\left(x,y,\frac{H_{\text{m}}}{2}\right)= 
h_{\text{c}}\widetilde{T}_{\text{m};2,2}\left(x,y,\frac{H_{\text{m}}}{2}\right);\nonumber
\end{gather}
\begin{gather}
-\kappa_{\text{m}}\frac{\partial}{\partial x}
\widetilde{T}_{\text{m};1}\left(\frac{W_{\text{m}}}{2},y,z,t\right)= 
h_{\text{c}}\widetilde{T}_{\text{m};1}\left(\frac{W_{\text{m}}}{2},y,z,t\right);\nonumber\\
\label{bwire_eq23}
-\kappa_{\text{m}}\frac{\partial}{\partial x}
\widetilde{T}_{\text{m};2,1}\left(\frac{W_{\text{m}}}{2},y,z\right)= 
h_{\text{c}}\widetilde{T}_{\text{m};2,1}\left(\frac{W_{\text{m}}}{2},y,z\right);\\
-\kappa_{\text{m}}\frac{\partial}{\partial x}
\widetilde{T}_{\text{m};2,2}\left(\frac{W_{\text{m}}}{2},y,z\right)= 
h_{\text{c}}\widetilde{T}_{\text{m};2,2}\left(\frac{W_{\text{m}}}{2},y,z\right);\nonumber
\end{gather}
\begin{alignat}{2}
\widetilde{T}_{\text{m};1}\left(x,0,z,t\right)&=0;&\,
-\kappa_{\text{m}}\frac{\partial}{\partial y}
\widetilde{T}_{\text{m};1}\left(x,L_{\text{w}},z,t\right)&=0;\nonumber\\ 
\label{bwire_eq24}
\widetilde{T}_{\text{m};2,1}\left(x,0,z\right)&=T_{\text{ch}}-T_{0};&\;
-\kappa_{\text{m}}\frac{\partial}{\partial y}
\widetilde{T}_{\text{m};2,1}\left(x,L_{\text{w}},z\right)&=0;\\
\widetilde{T}_{\text{m};2,2}\left(x,0,z\right)&=0;&\;
-\kappa_{\text{m}}\frac{\partial}{\partial y}
\widetilde{T}_{\text{m};2,2}\left(x,L_{\text{w}},z\right)&=0;\nonumber
\end{alignat}
\begin{alignat}{2}
\frac{\partial}{\partial x}\widetilde{T}_{\text{m};1}\left(0,y,z,t\right) &=0;&\;
\widetilde{T}_{\text{m};1}\left(x,y,-\frac{H_{\text{m}}}{2},t\right) &=0;\nonumber\\ 
\label{bwire_eq25}
\frac{\partial}{\partial x}\widetilde{T}_{\text{m};2,1}\left(0,y,z\right) &=0;&\;
\widetilde{T}_{\text{m};2,1}\left(x,y,-\frac{H_{\text{m}}}{2}\right) &=0;\\
\frac{\partial}{\partial x}\widetilde{T}_{\text{m};2,2}\left(0,y,z\right) &=0;&\;
\widetilde{T}_{\text{m};2,2}\left(x,y,-\frac{H_{\text{m}}}{2}\right) &=T_{\text{d}}-T_{0}.\nonumber
\end{alignat} Thus, by applying the method of separation of variables,
we arrive at
\begin{multline}
\label{bwire_eq26}
\widetilde{T}_{\text{m};1}\!=\!
\sum_{n}\sum_{m}\sum_{p}C^{\text{t}}_{\text{m};n,m,p}e^{-\frac{\kappa_{\text{m}}}{\rho_{\text{m}}c_{\text{e};\text{m}}}
\left(\lambda^{2}_{x;\text{m},n}+\lambda^{2}_{y;\text{m},m}+
\lambda^{2}_{z;\text{m},p}\right)t}\\
\cos\left(\lambda_{x;\text{m},n}x\right)\sin\left(\lambda_{y;\text{m},m}y\right)
\sin\left(\lambda_{z;\text{m},p}\left(z+\frac{H_{\text{m}}}{2}\right)\right),
\end{multline}
\begin{multline}
\label{bwire_eq27}
\widetilde{T}_{\text{m};2,1}\!=\!
\sum_{n}\sum_{p}C^{\text{s}}_{\text{m};1;n,p}e^{\lambda_{y;\text{m};n,p}y}
\left(1+e^{2\lambda_{y;\text{m};n,p}\left(L_{\text{w}}-y\right)}\right)\\
\cos\left(\lambda_{x;\text{m},n}x\right)
\sin\left(\lambda_{z;\text{m},p}\left(z+\frac{H_{\text{m}}}{2}\right)\right),
\end{multline}
\begin{multline}
\label{bwire_eq28}
\widetilde{T}_{\text{m};2,2}\!=\!
\sum_{n}\sum_{m}C^{\text{s}}_{\text{m};2;n,m}
\left(\!e^{\lambda_{z;\text{m};n,m}z}\!+\!
\frac{\left(h_{\text{c}}+\kappa_{\text{m}}\lambda_{z;\text{m};n,m}\right)}
{\left(\kappa_{\text{m}}\lambda_{z;\text{m};n,m}-h_{\text{c}}\right)}\right.\\
\left.e^{-\lambda_{z;\text{m};n,m}\left(z-H_{\text{m}}\right)}\right)
\cos\left(\lambda_{x;\text{m},n}x\right)\sin\left(\lambda_{y;\text{m},m}y\right),
\end{multline} where $\lambda_{x;\text{m},n}$, $\lambda_{y;\text{m},m}$, and 
$\lambda_{z;\text{m},p}$ are solutions of the 
characteristic equations 
\begin{alignat*}{2}
\lambda_{x;\text{m},n}\tan\left(\lambda_{x;\text{m},n}\frac{W_{\text{m}}}{2}\right)&=\!\frac{h_{\text{c}}}{\kappa_{\text{m}}},&\;
\lambda_{z;\text{m},p}\cot\left(\lambda_{z;\text{m},p}H_{\text{m}}\right)&=\!-\frac{h_{\text{c}}}{\kappa_{\text{m}}},
\end{alignat*}
\begin{equation}
\label{bwire_eq29}
\lambda_{y;\text{m},m}=\frac{\left(2m+1\right)\pi}{2L_{\text{w}}},
\;m\geq 0;
\end{equation} whereas $\lambda_{y;\text{m},n,p}$, and 
$\lambda_{z;\text{m},n,m}$ are given by
\begin{alignat}{2}
\label{bwire_eq30}
\lambda^{2}_{y;\text{m},n,p}&=\lambda^{2}_{x;\text{m},n}+\lambda^{2}_{z;\text{m},p},&\;  
\lambda^{2}_{z;\text{m},n,m}&=\lambda^{2}_{x;\text{m},n}+\lambda^{2}_{y;\text{m},m}.
\end{alignat} Finally, the coefficients 
$\left\lbrace C^{\text{t}}_{\text{m};n,m,p}, C^{\text{s}}_{\text{m};1;n,p}, C^{\text{s}}_{\text{m};2;n,m}\right\rbrace$
are determined by the initial condition at $t=0$ and the relevant 
BCs in \eqref{bwire_eq24} and \eqref{bwire_eq25}.

The compound heat kernel $G_{\text{m}}$ is calculated similarly.
That is, we define $\widetilde{G}_{\text{m}}\equiv G_{\text{m}}-T_{0}$,
and expand $\widetilde{G}_{\text{m}}=\widetilde{G}_{\text{m};1}+
\widetilde{G}_{\text{m};2,1}+\widetilde{G}_{\text{m};2,2}$, and thus
by recalling that $G_{\text{m}}$ for $t>0$ satisfies   
\begin{equation}
\label{bwire_eq31} 
\rho_{\text{m}}c_{\text{e;m}}
\frac{\partial}{\partial t}G_{\text{m}}\left(x,y,z,t,y^{\prime}\right)=
\kappa_{\text{m}}\nabla^{2}G_{\text{m}}\left(x,y,z,t,y^{\prime}\right),
\end{equation} it is not difficult to realise that the components 
$\widetilde{G}_{\text{m};1}$, $\widetilde{G}_{\text{m};2,1}$, and 
$\widetilde{G}_{\text{m};2,2}$ adopt expressions similar to those 
in \eqref{bwire_eq26}--\eqref{bwire_eq28} with 
coefficients $\left\lbrace C^{\text{t}}_{\text{g};n,m,p},
C^{\text{s}}_{\text{g};1;n,p},C^{\text{s}}_{\text{g};2;n,m}\right\rbrace$,
respectively. These coefficients are also obtained as before; for instance,
$C^{\text{s}}_{\text{g};1;n,p}$ and $C^{\text{s}}_{\text{g};2;n,m}$ are
obtained by replacing $T_{\text{ch}}-T_{0}$ and $T_{\text{d}}-T_{0}$
with $-T_{0}$ in the corresponding BCs of \eqref{bwire_eq24} and \eqref{bwire_eq25},
while the transient coefficient $C^{\text{t}}_{\text{g};n,m,p}$ is obtained by
considering \eqref{bwire_eq31} around $t=0$ where 
the impulse source $\dot{q}_{\text{i}}=
\delta(x)\delta(y-y^{\prime})\delta(z)\delta(t)$ is defined.  
Therefore, it is the constant $C^{\text{t}}_{\text{g};n,m,p}$
that carries the dependence on $y^{\prime}$ associated with the 
Green's function.

\section{Determination of the Model Auxiliary Constants}
\label{bwire_appB}
The constants $\widetilde{T}_{\text{w;e}}$ and $\chi_{\text{w}}$  
are introduced in the model as consequence of the 
adopted linearisation. Strictly speaking,
the actual wire \emph{effective} temperature is the
average of the wire exact temperature distribution. 
In the model, however, this value
has become an auxiliary source term 
together with the wire-compound \emph{effective}
heat transfer coefficient $\chi_{\text{w}}$.

For starters, we realise that these auxiliary two constants
do not allow us, as inferred from \eqref{bwire_eq18} and \eqref{bwire_eq20},
to impose rigorously $T_{\text{w}}=
T_{\text{m}};\,\forall y,\,\forall t$ at the
common interface (see Fig.~\ref{bwire_fig2}),
and yet by introducing them, we facilitate
accounting for the thermal interaction
between the compound and the wire.
Having said this, we are still allowed 
to state 
\begin{equation}
\label{bwire_eq32}
\int\limits_{0}^{t_{\text{p}}}\int\limits_{0}^{L_{\text{w}}}      
\lim_{z\rightarrow 0}\lim_{x\rightarrow 0}
\widetilde{T}_{\text{m}}\left(x,y,z,t\right)\diff y\diff t=
\int\limits_{0}^{t_{\text{p}}}\int\limits_{0}^{L_{\text{w}}}
\widetilde{T}_{\text{w}}\left(y,t\right)\diff y\diff t.
\end{equation} along the wire-compound interface. 
Such a condition holds
true always, and it enables us to set an equation 
for computing $\widetilde{T}_{\text{w;e}}$ and
$\chi_{\text{w}}$ in an iterative manner.
Namely, at start we compute an \emph{effective}
wire temperature $\widetilde{T}^{(0)}_{\text{w;e}}$ 
as if there were no compound. Subsequently, a value  
$\chi^{(0)}_{\text{w}}$ is obtained from \eqref{bwire_eq32} 
by plugging $\widetilde{T}^{(0)}_{\text{w;e}}$ in 
\eqref{bwire_eq18} and \eqref{bwire_eq20}; these two
initial values of the constants are used to compute 
a new effective temperature $\widetilde{T}^{(1)}_{\text{w;e}}$
from \eqref{bwire_eq18}, which in turn is used to
derive $\chi^{(1)}_{\text{w}}$ from \eqref{bwire_eq32}
again. This procedure is repeated until 
the pair $\lbrace\widetilde{T}^{(i)}_{\text{w;e}}, \chi^{(i)}_{\text{w}}\rbrace$
stabilises. An important observation to keep in mind 
during the iterations is that the following constraint should hold
true for every computed pair $\lbrace\widetilde{T}^{(i)}_{\text{w;e}},\chi^{(i)}_{\text{w}}\rbrace$   
\begin{equation}
\label{bwire_eq46}
\widetilde{T}^{(i)}_{\text{w;e}}<\frac{2I^{2}_{0}\uprho_{\text{e;0}}\alpha_{\uprho}}
{\epsilon_{\text{w}}\upsigma\chi^{(i)}_{\text{w}}A_{\text{w}}C_{\text{w}}\lvert\alpha_{\kappa}\rvert}.    
\end{equation} This inequality stems from the very definition 
of these auxiliary constants and is readily obtained from \eqref{bwire_eq17}.  
\section*{Acknowledgment}
The work in this article was carried out within the framework 
of the \href{http://fp7-nanocops.eu/}{NanoCops project} and 
the authors would like to acknowledge its financial support.
The authors would also like to thank the Czech Ministry of Education 
for its financial support through the National Sustainability program 
under grant LO1401. During this research the infrastructure of the SIX
Center was used. Special thanks goes as well to 
Jiri Petrzela and Roman Sotner of the Brno University of 
Technology for conducting some of the
measurements retrieving the experimental 
data used in the optimisation.

\input{bondwire_ref}

\end{document}